\documentclass[12pt]{article}
\pdfoutput=1

\usepackage[nosort]{cite}
\usepackage{cite}
\usepackage{amsmath,pifont,bbding}
\usepackage{epsfig}
\usepackage{amsfonts}
\usepackage{amssymb}
\usepackage{multirow}
\usepackage{enumerate}
\usepackage{subfigure}
\usepackage{slashed}

\usepackage{color}

\usepackage{subfigure}

\usepackage{caption}

\newcommand{\be}{\begin{equation}}
\newcommand{\ee}{\end{equation}}
\newcommand{\bee}{\begin{equation*}}
\newcommand{\eee}{\end{equation*}}
\newcommand{\bea}{\begin{eqnarray}}
\newcommand{\eea}{\end{eqnarray}}
\newcommand{\bean}{\begin{eqnarray*}}
\newcommand{\eean}{\end{eqnarray*}}

\begin{document}

\setcounter{page}{0}
\thispagestyle{empty}

\begin{flushright}
CERN-PH-TH/2012-368\\
UCI-HEP-TR-2013-02\\
\today
\end{flushright}

\vskip 8pt

\begin{center}
{\bf \LARGE {
Gamma-ray lines and
One-Loop Continuum \\
\vskip 8pt
from $s$-channel Dark Matter Annihilations \\
 }}
\end{center}

\vskip 12pt

\begin{center}
 {\bf C.~B.~Jackson$^{a}$,  G\'eraldine  Servant$^{b,c,d}$, Gabe Shaughnessy$^{e}$, \\
 \vskip 4pt
 Tim M.P. Tait$^{f}$, and Marco Taoso$^{d,g}$ }
\end{center}

\vskip 14pt

\begin{center}

\centerline{$^{a}${\it University of Texas at Arlington, Arlington, TX 76019 USA}}
\centerline{$^{b}${\it CERN Physics Department, Theory Division, CH-1211 
Geneva 23, Switzerland}}
\centerline{$^{c}${\it ICREA at IFAE, Universitat Aut\`onoma de Barcelona, 08193 Bellaterra, Barcelona, Spain}}
\centerline{$^{d}${\it Institut de Physique Th\'eorique, CEA/Saclay, F-91191 
Gif-sur-Yvette C\'edex, France}}
\centerline{$^{e}${\it Department of Physics, University of Wisconsin, Madison, WI 53706 USA}}
\centerline{$^{f}${\it Department of Physics \& Astronomy, University of California, Irvine, CA 92697 USA}}
\centerline{$^{g}${\it Department of Physics \& Astronomy, University of British Columbia,
Vancouver, BC V6T 1Z1 Canada}}
\vskip .3cm
\centerline{\tt geraldine.servant@cern.ch, chris@uta.edu,}
\centerline{\tt gshau@hep.wisc.edu, ttait@uci.edu, marco.taoso@cea.fr}
\end{center}

\vskip 10pt

\begin{abstract}
\vskip 3pt
\noindent

The era of indirect detection searches for dark matter has begun, with
the sensitivities of  gamma-ray 
detectors now approaching the parameter space relevant for weakly interacting 
massive particles. 
In particular, gamma ray lines would be 
smoking gun signatures of dark matter annihilation,
although they are typically suppressed compared to the continuum.
In this paper, we pay particular attention to the 1-loop continuum generated together 
with the gamma-ray lines and  investigate under which conditions a dark matter model 
can naturally lead to a line signal that is  relatively enhanced.
We study generic classes of models in which DM is a fermion that annihilates 
through an $s$-channel mediator which is either a vector or scalar 
and identify the coupling and mass conditions under which large line signals occur.  
We focus on the ``forbidden channel mechanism" 
advocated a few years ago in the ``Higgs in space" 
scenario for which tree level annihilation is kinematically forbidden today. 
Detailed calculations of all 1-loop annihilation channels are provided. We single out 
very simple models with a large line over continuum ratio 
and present general predictions for a large range of WIMP masses that are relevant 
not only for Fermi and Hess II but also 
for the next generation of telescopes such as CTA and Gamma-400.
Constraints from the relic abundance, direct detection and collider bounds are also 
discussed.

\end{abstract}

\newpage


\vskip 13pt

\section{Introduction}

If dark matter (DM) is a thermally produced Weakly Interacting Massive Particle
(WIMP), there is a lower bound on its total 
annihilation cross section to avoid overabundance of DM in the universe:
independently of the details of thermal freeze-out and 
whether DM is ``symmetric" or not, its thermally averaged annihilation cross section  
at the time of freeze-out $\langle \sigma v\rangle$ should be greater than 
order 1 pb~$\sim 10^{-26}$ cm$^3$/s.  
Assuming one saturates this bound (or in other words that the dark matter is
a thermal relic), one can classify the prospects for the indirect detection of
dark matter based on whether the annihilation is $s$-wave, 
$\langle \sigma v\rangle \propto 1$, or $p$-wave, 
$\langle \sigma v\rangle \propto v^2$, where $v$ is the relative velocity of the
annihilating particles.
For a $p$-wave suppressed cross section, the prospects of experimentally
observing annihilation products are poor, since $v$ has fallen
from ${\cal O}(1)$ at the time of freeze-out to $\sim 10^{-3}$ in our galaxy today. 
On the other hand, an $s$-wave annihilator has essentially 
the same $\langle \sigma v\rangle$ today as in the early universe, and indirect 
detection experiments are bearing down on the correct range.  In fact, 
for WIMP masses below $\sim 50$ GeV, the thermal freeze-out cross section 
is narrowly excluded by null results from the Fermi LAT 
\cite{Ackermann:2011wa,Ackermann:2012qk}.

In this work, we 
explore a class of models where annihilations in the early universe and annihilations 
today are controlled by different processes and thus are naturally of
different sizes.
Such a situation is motivated in models where the DM 
couples most significantly to particles with heavy masses.
For instance, in scenarios of composite Higgs and top quarks, the new 
physics sector responsible for EW symmetry breaking only has sizable couplings with
heavy states. If the DM is part of this sector, 
annihilation into light Standard Model (SM) states is often suppressed.
If the mass difference between DM and the new heavy particles (which
we generically denote as $\psi$) is less than a 
few tens of GeV, the DM kinetic energy in the early universe allows tree level 
annihilation into $\psi$ pairs and the correct relic abundance can be obtained for 
couplings of order 1. On the other hand, in this ``forbidden channel" scenario,
tree level annihilation into $\psi$ pairs is 
kinematically forbidden in the current epoch\footnote{
Another interesting aspect of this class of models is that direct detection constraints are often naturally eluded.}.

We proposed such a scenario a few years ago in the ``Higgs in Space" model
\cite{Jackson:2009kg}, which was inspired by models of (partial) 
compositeness where the top quark is the only SM particle with sizable couplings to 
the new physics sector \cite{Lillie:2007hd,Kumar:2009vs,Pomarol:2008bh,Agashe:2004bm}.
In a simple realization we explored a $Z^\prime$ vector
mediator which had large couplings to DM and to the top quark but suppressed 
couplings to the other SM particles.  Thermal freeze-out in the early universe is
controlled by DM annihilation into top pairs, and for DM masses slightly below
the top mass ($\sim 150$~GeV), one can obtain the correct relic abundance while 
having very suppressed annihilation into the three-body final state $t W b$
in our galaxy.  Annihilation can be dominated by one loop channels, and this
scenario can naturally lead to large gamma ray line signatures
relative to the continuum.

\begin{figure}[t!]
\begin{center}
\includegraphics[angle=0,width=0.45\linewidth]{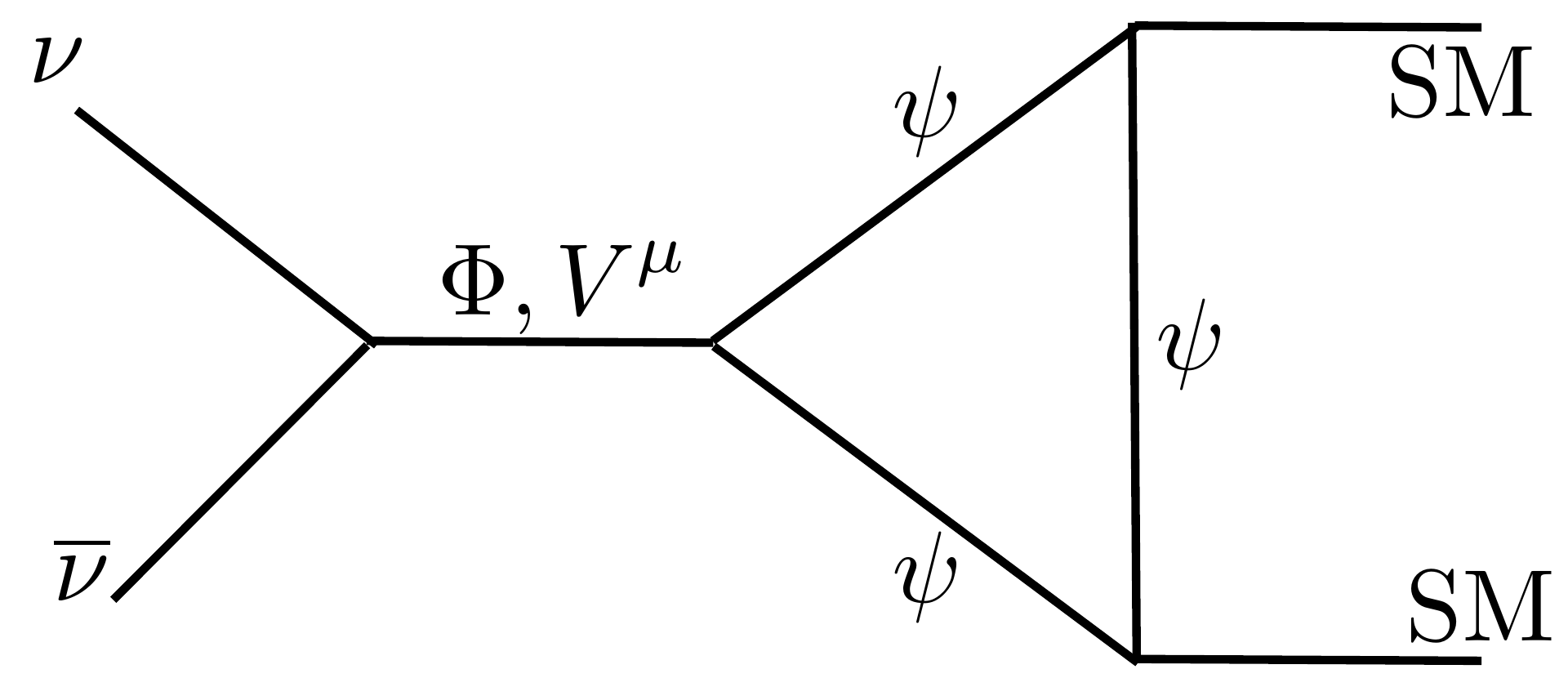}
\caption{\small General topology studied in this paper. DM ($\nu$) annihilates through an $s$-channel scalar $\Phi$ or vector $V^{\mu}$  resonance and loop of new fermions $\psi$.}
\label{fig:topology} 
\end{center}
\end{figure}

In this work, we consider a larger class of scenarios in
the same category.  In all of the theories, we assume the DM is a Dirac fermion
whose primary interactions with the SM are through a scalar or vector
$s$-channel mediator, leading to annihilations of the form shown in
Fig.~\ref{fig:topology}.  While there has recently been considerable interest in
similar theories \cite{Buckley:2012ws}, 
we flesh out the discussion in a number of important ways. 
In particular, we demonstrate that it is not sufficient to
suppress tree level annihilations; one must also take a close look
at the one-loop contributions to the continuum gamma rays as well.  We carry out
a thorough investigation of this question by studying representative examples.

The plan is the following: in Section \ref{sec:scalar} we illustrate the idea with 
a very simple model where DM annihilates via a scalar $\Phi$ into
a heavy fermion $\psi$.  We explore different choices of couplings and
gauge representations for $\psi$.
We compute in detail all one-loop annihilation channels and determine the line-to-
continuum ratio $\sigma{\mbox{\tiny line}}/\sigma{\mbox{\tiny continuum}}$,
which characterizes the strength of the line signal relative to the DM-induced
continuum background, and provides a useful analysis handle on any putative
line signal \cite{Cohen:2012me}.
Section \ref{sec:vector} discusses the vector mediator case, distinguishing vector and 
axial couplings. If DM has axial couplings, both vector and scalar $s$-channel 
exchanges can play an important role. 
We compute all one-loop contributions to the continuum for various assumptions on 
the nature of the couplings (scalar, pseudoscalar, vector, axial).
 We summarize our 
conclusions in Section \ref{sec:summary}.
In the appendices, 
we collect the expressions describing all of the one-loop annihilation 
processes.  These results go somewhat beyond the immediate needs of the
current work, and are easily adapted to any WIMP model whose interactions with
the SM are primarily through an $s$-channel mediator.

\section{Scalar Mediator Model}
\label{sec:scalar}

As a first example illustrating the forbidden channel scenario, 
we consider a model with a minimal dark sector:
in addition to the (SM singlet) Dirac fermion DM, $\nu$, we include
another Dirac fermion $\psi$ with vector-like SM gauge interactions which
acts as the portal to the SM.  
$\nu$ and $\psi$ can interact by exchanging a real scalar singlet $\Phi$. 
We will eventually consider various possibilities for the gauge quantum numbers of 
$\psi$, but begin by assuming that it is
is a color singlet and a doublet under $SU(2)$ carrying the opposite hypercharge $Y=1/2$ to the SM lepton doublet.
In that case, the relevant interactions are given by
 \be
{\cal L} \supset - \overline{\nu} (y_{\nu \Phi}^S + iy_{\nu \Phi}^P \gamma^5) \nu \Phi
- \overline{\psi} (y_{\psi \Phi}^S + iy_{\psi \Phi}^P \gamma^5) \psi \Phi
- y_H ~ \overline{\psi} H \nu + h.c
\label{eq:Lscalar}
\ee
where $H$ is the SM Higgs doublet and we have assumed as usual
that a dark symmetry (parity or $U(1)_D$) forbids
$\nu$ from mixing with the SM neutrinos via interactions such as $L H \nu$.  The
last term is responsible for the decay of $\psi \rightarrow h \nu$
(where $h$ is the SM Higgs boson), with 
$y_H$ chosen to be smaller than $\sim 10^{-3}$ to avoid a degree of
$\psi^0-\nu$ mixing which induces a dangerous $Z$-$\nu$-$\bar{\nu}$ coupling excluded by 
direct DM searches, but large enough for $\psi$ to decay before BBN.  
For this 
range of values, it is safely small enough as to be roughly irrelevant for the 
remainder of our discussion.

In addition, there are interactions such as $|H|^2 \Phi$ which after electroweak 
breaking induce mixing between $\Phi$ and the SM Higgs.  We characterize
the degree of mixing as $\sin \alpha$, with the unmixed limit obtained
as $\sin \alpha \rightarrow 0$.  A small amount of
such mixing is welcome to allow $\Phi$ to decay, and as we shall see below
can help arrange for the correct $\nu$ relic density, but a large degree of
mixing would lead to a large direct detection rate which could be problematic.
We will discuss ranges of $\sin \alpha$ together with the relic density in more detail,
below.

%
\begin{figure}[t!]
\begin{center}
\includegraphics[angle=0,width=0.5\linewidth]{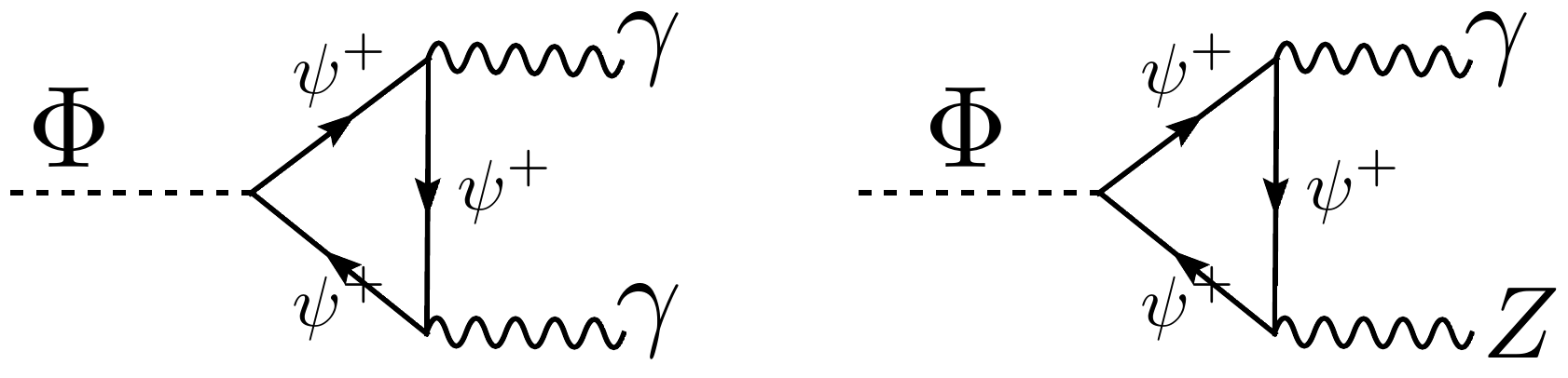}
\includegraphics[angle=0,width=0.99\linewidth]{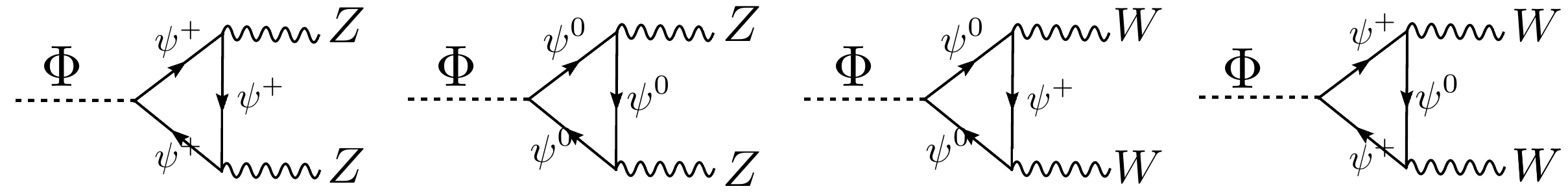}
\includegraphics[angle=0,width=0.5\linewidth]{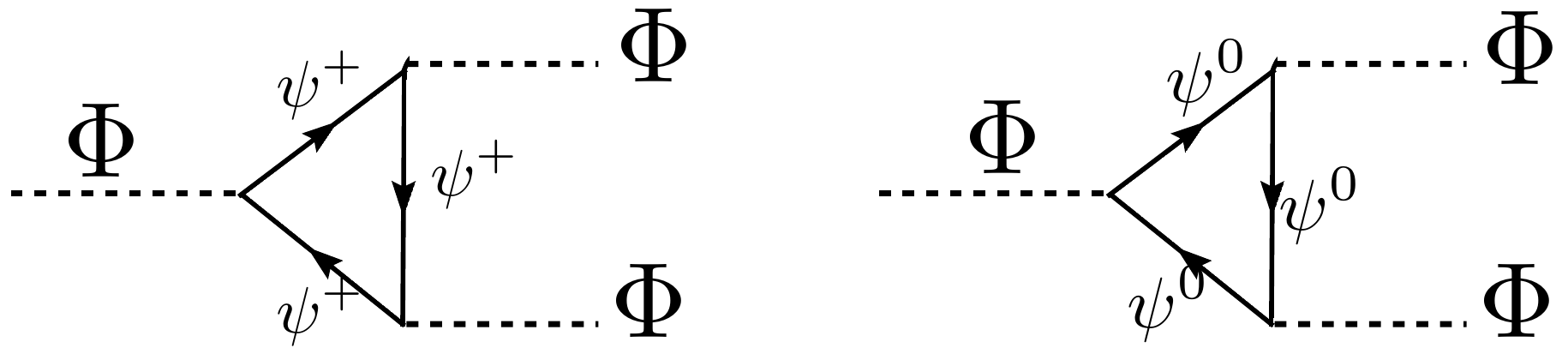}\\
\hspace{1.2cm} \includegraphics[angle=0,width=0.6\linewidth]{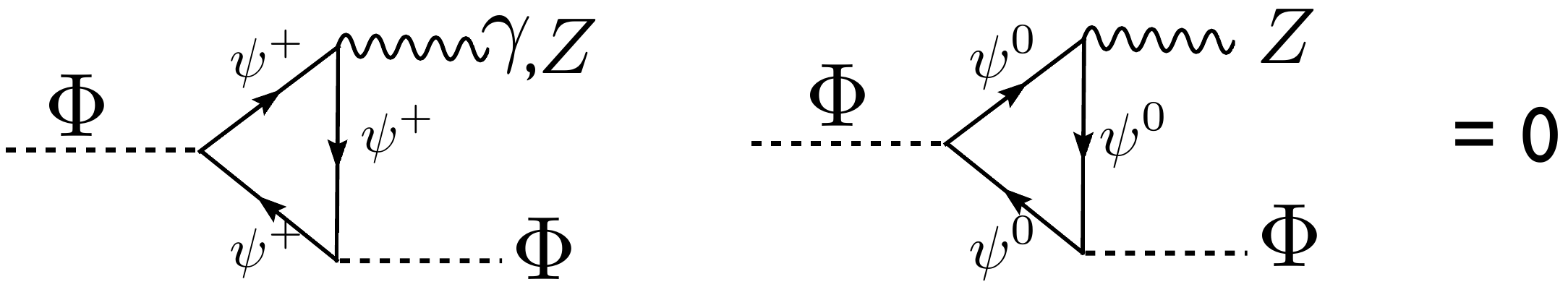}
\end{center}
\caption{Representative Feynman diagrams for relevant one loop processes 
in the scalar model with an $SU(2)$ doublet $\psi$ with $Y=1/2$.  The last two diagrams,
corresponding to annihilation into $\gamma \Phi$ and $Z \Phi$, vanish due to 
angular momentum and CP conservation respectively.}
\label{fig:scalar_model2} 
\end{figure}

The one-loop annihilation channels mediated by $\Phi$ are shown in 
Fig.~\ref{fig:scalar_model2}, including potentially three
lines (corresponding to $\gamma \gamma$, $\gamma Z$, and
$\gamma \Phi$) at distinct energies\footnote{The presence of
a forest of multiple lines is a generic feature arising from
$SU(2) \times U(1)$ gauge 
invariance \cite{Bertone:2009cb,Rajaraman:2012db}.}.  
However, in practice the $\gamma \Phi$ (and $Z \Phi$) channels
mediated by an $s$-channel scalar vanish because they do not conserve
angular momentum or CP.
All of
the corresponding cross sections scale as  
\be
\langle \sigma v \rangle \propto 
\left(  (y^S_{\nu \Phi})^2 v^2 + 4 (y^P_{\nu \Phi})^2    \right)
\ee
which in the non-relativistic limit is $p$-wave suppressed for purely scalar couplings. 
We therefore concentrate on pseudo-scalar couplings of $\Phi$ to DM to be able to 
observe a sizable $\gamma$-ray line signal (in which case the elastic scattering via 
Higgs exchange after $h-\Phi$ mixing is velocity suppressed). 
On the other hand, large line signals are not predicated on the nature of the 
$\Phi-\psi$ coupling.  
Detailed expressions in the $v \to 0$ limit are given in the appendices.
In addition, the final states consisting of a vector and a scalar particle
vanish for the $s$-channel scalar exchange.
A summary of the various 
channels for the scalar (S) and pseudo-scalar (P) $\Phi-\psi$ couplings is shown in 
Table~\ref{tab:unmixedPhi}.
It is worth noting that the $t$-channel tree level annihilation into $\Phi \Phi$ is 
$p$-wave suppressed in all cases, and therefore annihilation into $\Phi \Phi$ 
is controlled by the one-loop contribution.

\begin{table}[t]
\begin{center}
\begin{tabular}{|c|c|c|}
\hline
&&\\
$y_{\psi \Phi}$&$S$ &$P$ \\
\hline 
$\gamma \gamma$ & \Checkmark&  \Checkmark\\
\hline
$\gamma Z$ & \Checkmark&  \Checkmark\\
\hline
$\gamma \Phi$  & 0&0\\
\hline
$Z \Phi$  & 0&0\\
\hline
$\Phi \Phi$  & \Checkmark&0\\
\hline
$ZZ$ & \Checkmark&  \Checkmark\\
\hline
$WW$  & \Checkmark& \Checkmark\\
\hline
\end{tabular}
\hspace{1.2cm}
\begin{tabular}{|c|c|c|c|}
\hline
&$\psi_{-1}$&$\psi_{-3/2}$&$\psi_{1/2}$\\
\hline
$\sigma_{\gamma Z}/\sigma_{cont}$  &10&0.5&0.02 \\
\hline
$\sigma_{\gamma \gamma}/\sigma_{cont}$ &30 &1&0.04 \\
\hline
\end{tabular}
\end{center}
\caption{Summary of the one-loop DM annihilation channels in scalar
exchange models and typical (but parameter-dependent) values of the line/continuum ratio for the three 
charge assignments discussed in the text. $\sigma_{cont}$ is the sum of all the 1-loop continuum cross sections. } 
\label{tab:unmixedPhi}
\end{table}

As alluded to above, there is a wider variety of SM $SU(2) \times U(1)$ gauge
charges for $\psi$ which can result in a viable line signal.  We study the particular
cases of SM representations ($SU(3)$, $SU(2)$, $U(1)$):
 \begin{itemize}
 \item (1,~2,~1/2): $\psi_{1/2}=(\psi^+,\psi^0)$;
 \item   (1,~2,~-3/2): $\psi_{-3/2}=(\psi^{-},\psi^{--})$;
 \item  (1,~1,~-1): $\psi_{-1}=\psi^-$.
 \end{itemize}   
In terms of the dark symmetry, in the case $\psi=\psi_{1/2}$, 
$\psi$ is charged under the dark symmetry as well and decays into $\nu H$;
when $\psi=\psi_{-3/2}$, $\psi$ is uncharged and can decay into $H e_R$; 
and when $\psi=\psi_{-1}$, we also assume it is uncharged and decays into SM states 
via a tiny mixing with the SM right-handed lepton (note that 
the last term in Eq.~(\ref{eq:Lscalar}) is only present for $\psi = \psi_{1/2}$ ).
In all cases, these couplings allowing $\psi$ to decay can
be chosen small enough that they do not play much role in dark matter searches
or in determining the relic density.

\begin{figure}[h!]
\begin{center}
\includegraphics[angle=0,width=0.71\linewidth]{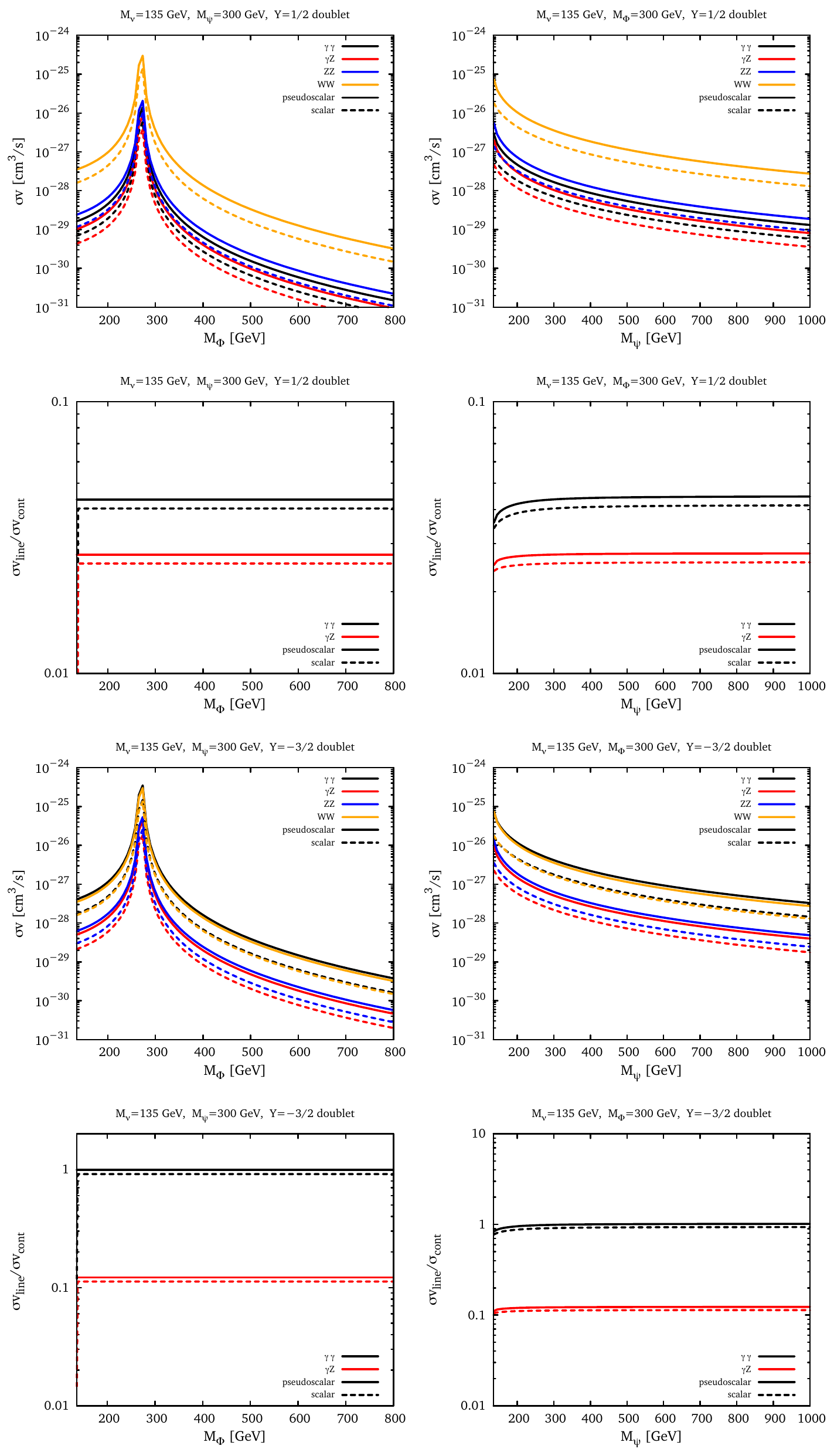}
\end{center}
\caption{ \small Gamma ray line signal and continuum strength for
scalar exchange models: top: the (1,~2,~1/2) case;
bottom: the (1,~2,~-3/2)  case.}
\label{fig:Ym32}
\end{figure}
 
\begin{figure}[t!]
\begin{center}
\includegraphics[angle=0,width=0.95\linewidth]{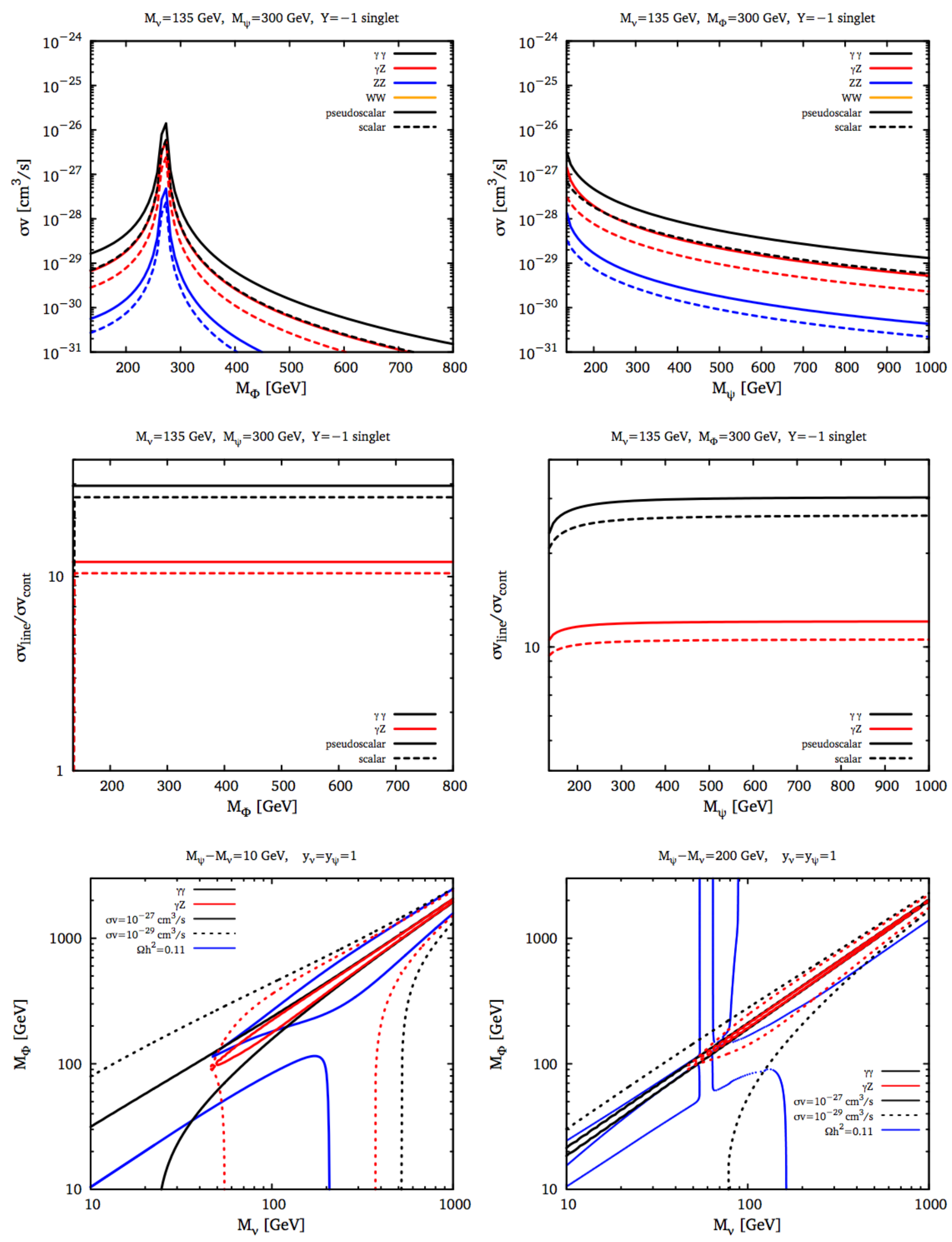}
\end{center}
\caption{\small Gamma ray line signal and continuum strength for
scalar exchange models in the (1,~1,~-1) case. In the last two plots, we added the contours corresponding to correct relic abundance.}
\label{fig:contoursscalar} 
\end{figure}
 
\begin{figure}[t!]
\begin{center}
\includegraphics[angle=0,width=0.326\linewidth]{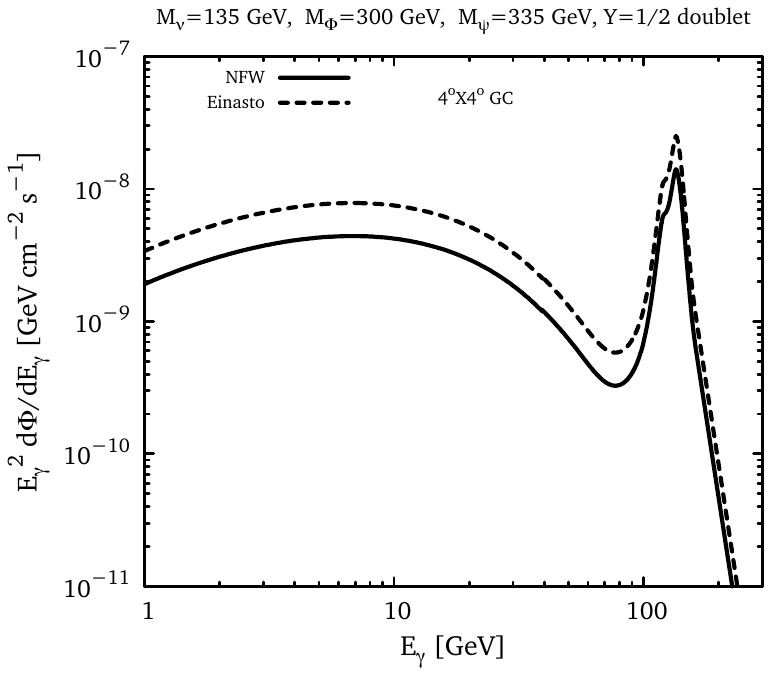}
\includegraphics[angle=0,width=0.326\linewidth]{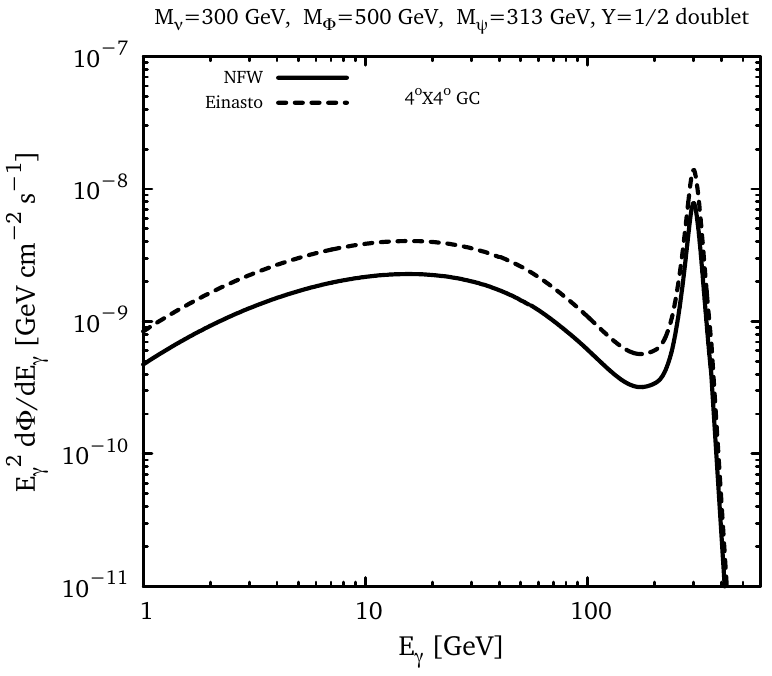}
\includegraphics[angle=0,width=0.332\linewidth]{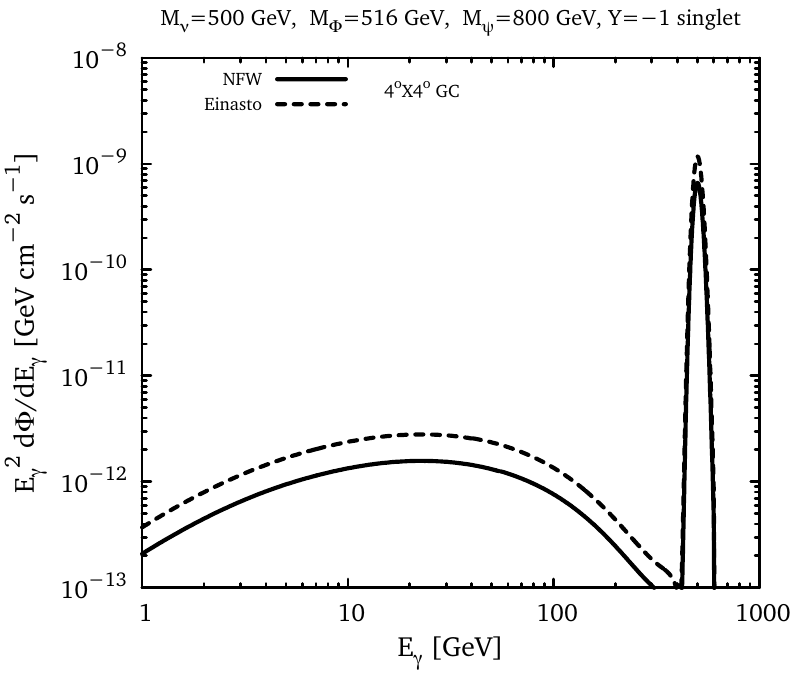}
\end{center}
\caption{\small Examples gamma-ray spectra for three choices of parameters leading 
to the correct DM relic abundance and for two choices of DM profiles.}
\label{fig:gamma-spect} 
\end{figure}

In Figures \ref{fig:Ym32} and \ref{fig:contoursscalar},  we show the (complete one
loop) cross sections for 
the cases where $\psi$ has scalar (dotted) or pseudo scalar (solid) coupling.
We illustrate the resonant dependence on $M_\Phi$ for fixed DM mass, the
dependence on $M_\psi$, the line cross section to continuum cross section,
$\sigma{\mbox{\tiny line}}/\sigma{\mbox{\tiny continuum}}$, and contours of
fixed cross sections the $M_\nu$-$M_\Phi$ parameter plane. 
For $\psi=\psi_{1/2}$, the $WW$ cross section is large relative to the 
$\gamma \gamma$ or $\gamma Z$ channels and the line to 
continuum ratio $\sigma{\mbox{\tiny line}}/\sigma{\mbox{\tiny continuum}}$ is typically 
around 0.02 to 0.04 (still larger than what a random WIMP model with tree 
level annihilation would predict \cite{Cohen:2012me}). 
Much larger ratios can be obtained  for the two other cases. 
For $\psi=\psi_{-3/2}$, the ratio goes up to $\sim 1$, with 
both $\psi^{-}$ and $\psi^{--}$ contributing to the line signal.
When $\psi$ is an $SU(2)$ singlet, $\psi=\psi_{-1}$, there is no annihilation into
$WW$, with the line to continuum ratio jumping to values as large as $\sim 30$.
We illustrate the spectrum of a few promising cases in Figure~\ref{fig:gamma-spect}.

Since all of the channels experience the same resonant enhancement, 
the ratios do not depend on $M_{\Phi}$ (though as shown, together with
the DM mass, it does control the absolute values of the cross sections). 
Increasing $M_{\psi}$ leads to the fall-off in the absolute rates shown on the figures. 
We have fixed the DM mass to 135 GeV for illustration, inspired by the
features in the Fermi LAT data from the galactic center
\cite{Rajaraman:2012db,Weniger:2012tx,Whiteson:2012hr}; 
the values of the ratios are not very sensitive to this choice of mass and one
can easily realize strong lines at other energies. 

\begin{figure}[t!]
\begin{center}
\includegraphics[angle=0,width=0.69\linewidth]{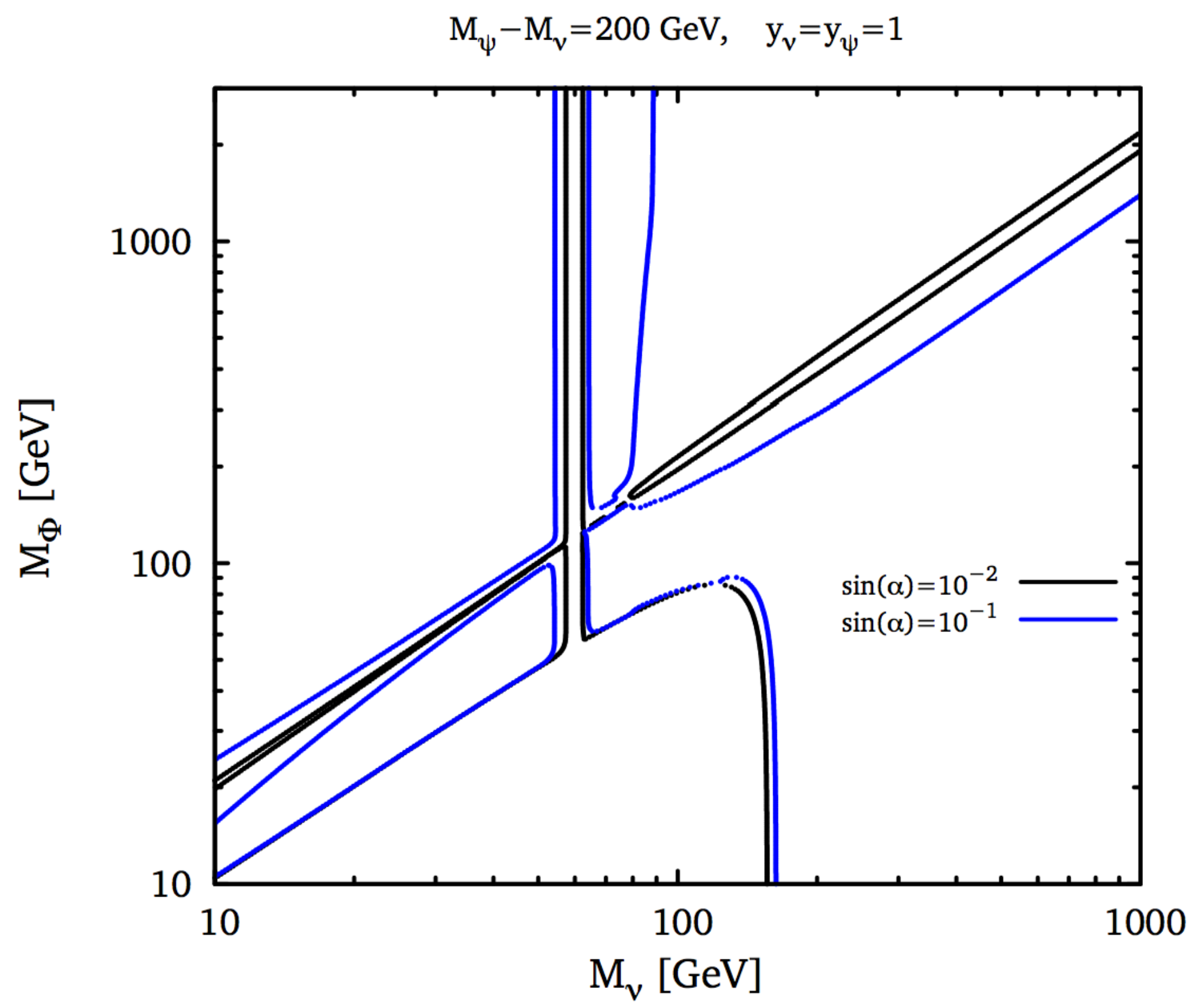}
\end{center}
\caption{\small Contours of correct relic abundance in the 
$M_{\nu}$-$M_{\Phi}$ plane from annihilations into $\psi \psi$ and $\Phi \Phi$ (where allowed) 
and SM states via $h-\Phi$ mixing. }
\label{fig:reliccontour} 
\end{figure}

This class of models provides simple examples of potentially large gamma 
ray line signatures.  It is also easy to obtain the correct relic abundance in this setup.
For non-zero $\sin \alpha$, the $h$-$\Phi$ mixing produces resonant enhancement
of the annihilation cross section for $M_\nu \simeq M_h / 2 \simeq 62$~GeV,
and for $M_\nu \simeq M_\Phi / 2$.  The width of $\Phi$ and its coupling to SM
final states are both controlled by $\sin \alpha$.  The resulting relic density is
a complicated function of the DM mass, as illustrated in Fig.~\ref{fig:reliccontour}.
Typically, there will be regions around both the $h$ and $\Phi$ resonances
with the correct relic abundance.  Obviously, the regions around the $\Phi$
resonance will also lead to large gamma ray line features.  Since the $\psi$ mass is
slightly above the DM mass, one could also invoke coannihilation between
$\nu$ and $\psi$.  
Similarly, when $\psi$ decays into SM states and it is slightly heavier than $\nu$,
$\nu \bar{\nu} \to \psi \bar{\psi}$ annihilations are efficient in the early universe, 
although forbidden today.  This is the forbidden channel idea proposed 
in \cite{Jackson:2009kg}, where the role of $\psi$ was played by the top quark.
 While it is generically true that one can arrange for both large gamma ray line
signals and the correct relic density, there is no firm generic correlation between the
two processes.

 We close this section with some brief discussion of collider and precision
 measurements.  Since all choices of $\psi$ are vector-like, with at most extremely
 tiny mixing with SM fermions, constraints to precision electroweak observables
 \cite{Peskin:1991sw}
 are expected to be much weaker than direct constraints from colliders.  Since the
 $\psi$ states contain charged particles which can decay into leptons, low-background
 final states at the LHC are possible; however, these are balanced by very small
electroweak production cross sections.  As a result, current 
experimental constraints are rather weak, and new states as light as a few hundreds 
of GeV are allowed.  We expect that further high energy running of the LHC should
eventually cover a good portion of the parameter space.

\section{Vector Mediator Model} 
\label{sec:vector}

The case of an $s$-channel 
vector mediator builds on what we have learned from the scalar model in the
previous section.  The SM gauge sector is extended by a $U(1)^\prime$ gauge
symmetry which is broken by the VEV of a complex
scalar, resulting in a spectrum consisting of a massive vector
boson $Z^\prime$ and the associated Higgs scalar $\Phi$.  Their masses
can be related to the $U(1)^\prime$-breaking VEV,
 \bea
M_{Z'}&=& g_{Z'} q_{\Phi}~ \langle \Phi \rangle , \\
M_{\Phi}&=& \sqrt{a}~ \langle \Phi \rangle  .
\eea
where $q_\Phi$ is the charge of the complex scalar, $g_{Z^\prime}$ is the
$U(1)^\prime$ gauge coupling, and $a$ is the scalar quartic, which
we will fix to 1 to illustrate the typical physics.  The phenomenology
associated with $\Phi$ exchange
is very similar to the scalar model of the previous section, including
the possibility that it mixes with the SM Higgs $h$ through couplings
such as $| H |^2 | \Phi |^2$.

The DM is charged under $U(1)^\prime$, as are the $\psi$ fermions.
As in the case of the scalar exchange models, there are a variety of
viable choices for the representations of $\psi$ under the SM.  
In practice, we choose $\psi$ to be an $SU(2)$ singlet carrying one
unit of hypercharge.
In principle, nothing determines the $SU(3)$ representation of
$\psi$; in practice we choose them
to be uncolored in order to evade large annihilation into $gg$
\cite{Chu:2012qy,Zhang:2012da}
as well as the LHC bound on $t^\prime$-like objects of
about 450 GeV \cite{Rao:2012gf,CMS:2012ab}.
We leave the case where $\psi$ is colored (and can mix) with
the right-handed top quark for separate work \cite{Jackson:2013tca}.

Any additional $U(1)^\prime$ symmetry will generally kinetically mix with the SM hyper-charge
boson.  This mixing is a free parameter of the theory, and is additively renormalized by loops of the
$\psi$ fields.  In order not to induce a large contribution to direct scattering, the mixing parameter
should be smaller than of order $10^{-3}$, consistent with a value induced by loop processes
without any large cancellations \cite{Jackson:2013tca}. 

\subsection{Mass Structure versus Couplings}

We will find that the one-loop cross sections in this setup are 
very sensitive to the vector or axial nature of  the couplings, as 
summarized in Table~\ref{tab:unmixedZp}. 
If all couplings in the loop are vector-like, there is no one-loop 
annihilation into two gauge bosons. In practice, an odd number of 
axial vector couplings is required in the loop for the 
annihilation cross section to survive in the non-relativistic limit.
To avoid SM anomalies, we take  $\psi$ to have vector-like couplings with  
respect to the SM gauge group,
therefore, strong line signals require axial vector couplings of
the $Z^\prime$ to $\nu$ or $\psi$ (or both). This implies that
both states are actually hybrids of underlying states with different $U(1)^\prime$
charges, married together by $\langle \Phi \rangle$.

 In order to avoid
mixed $U(1)^\prime$-SM anomalies, we engineer this by including vector-like
{\em pairs}
of $\nu$ with charges $q_{\nu_1}$ and $q_{\nu_2}= q_{\nu_1} - q_\Phi$ 
 and another
vector-like pair of $\psi$ fields with charges
$q_{\psi_1}$ and $q_{\psi_2}= q_{\psi_1} - q_\Phi$.  
In principle, the right- and left-chiral members of each pair could be married
by a vector-like
mass term, but we neglect these terms, which would induce
dependence on a mixing angle, for simplicity\footnote{As such terms are technically
natural, there is no fine-tuning associated with this choice.}.

These charge assignments insure that
Yukawa interactions with $\Phi$ are allowed, and after obtaining a VEV,
result in mass terms for the two states in each pair,
\bea
M_{\psi}=\lambda^S_{\psi} \langle \Phi \rangle &,&  M_{\psi'}=\lambda^S_{\psi'} \langle \Phi \rangle , \\
M_{\nu}= \lambda^S_{\nu} \langle \Phi \rangle&,& M_{\nu'}= \lambda^S_{\nu'} \langle \Phi \rangle .
\eea
where $\psi_R \equiv \psi_{2R}$, $\psi_L \equiv \psi_{1L}$, $\psi'_R \equiv \psi_{1R}$ and $\psi'_L \equiv \psi_{2L}$.
These masses are bounded by perturbativity of the couplings to be
less than about $\lesssim 4 \pi \langle \Phi \rangle$.  

This construction guarantees the presence of an axial-vector interaction.  For models with purely vector-like
interactions, we can choose $q_{\nu_1}= q_{\nu_2}$ and $q_{\psi_1}= q_{\psi_2}$, which forbids interactions
with $\Phi$, but allows us to write down a gauge invariant mass for the field of interest.  In that case, the heavier
element of the pair is superfluous.

The most important terms in the Lagrangian read:
\begin{eqnarray}
 {\cal L}&\supset& i \ \bar{\nu}_1 \slashed{D} \nu_1 + i \ \bar{\nu}_2\slashed{D} \nu_2 \  
 + \  i \ \bar{\psi}_1 \slashed{D} \psi_1 + \  i \ \bar{\psi}_2 \slashed{D} \psi_2 \\
  &-&  \  ( \overline{\nu}_1 (y_{\nu \Phi}^S + iy_{\nu \Phi}^P \gamma^5) 
  \nu_2 \Phi + h.c )
\ - \  (  \overline{\psi}_1 (y_{\psi \Phi}^S + iy_{\psi \Phi}^P \gamma^5) 
\psi_2 \Phi  + h.c )~,
\nonumber
\label{vectorlagrangian}
 \end{eqnarray}
 where
 \begin{eqnarray}
{D_{\mu}} \nu&=&\partial_{\mu} \nu - i g_{ Z'}  q_{\nu}\  Z'_{\mu} \  \nu ~, \\
 {D_{\mu}} \psi&= &\partial_{\mu} \psi - i g_{Z'}  q_{\psi}  Z'_{\mu} \  \psi \ 
 - i g_{Z}    Z_{\mu} \   \psi - i Q^{\psi}_e e    A_{\mu} \    \psi ~,
  \end{eqnarray}
for each  $\nu_{1,2}$ and $\psi_{1,2}$, resulting in equal vector couplings and opposite axial-vector couplings of the mass eigenstates $\psi$ and $\psi'$  to $Z'$ as:
\begin{eqnarray}
 {\cal L}&\supset& g_{Z'} (q_{\psi_1} -\frac{q_{\Phi}}{2}) \bar{\psi}\gamma^{\mu} Z^{\prime}_{\mu} \psi - \frac{g_{Z'}}{2}  \bar{\psi}\gamma^{\mu} \gamma^5Z^{\prime}_{\mu} \psi \\
 \nonumber
 &+& g_{Z'} (q_{\psi_1} -\frac{q_{\Phi}}{2}) \bar{\psi'}\gamma^{\mu} Z^{\prime}_{\mu} \psi' + \frac{g_{Z'}}{2}  \bar{\psi'}\gamma^{\mu} \gamma^5Z^{\prime}_{\mu} \psi' \\
\label{axiallagrangian}
 \end{eqnarray}

\begin{figure}[t!]
\begin{center}
\includegraphics[angle=0,width=0.92\linewidth]
{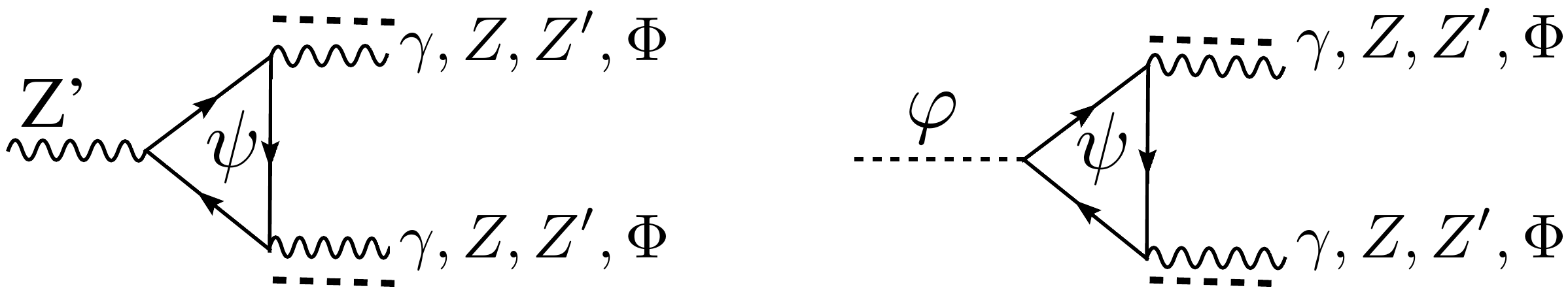}
\end{center}
\caption{ \small Representative one-loop annihilation channels in the vector model. 
The status of each channel depending on the nature of the couplings is summarized 
in Table \ref{tab:unmixedZp}.}
\label{fig:extras} 
\end{figure}

This module allows for tree level annihilation into 
$Z^\prime Z^\prime$, $Z^\prime \Phi$ and $\psi \overline{\psi}$, 
all of which would lead 
to a large gamma-ray continuum.  These potentially over-whelming
contributions may be suppressed by imposing
$M_{\nu} < M_{Z^\prime}$ (i.e. $\lambda^S_{\nu}< g_{Z^\prime}q_{\Phi}$),
$M_{\nu} < (M_{\Phi}+M_{Z^\prime})/2$, and $\lambda^S_{\nu}<\lambda^S_{\psi}$.
Annihilation into $\Phi \Phi$ is $p$-wave suppressed, and thus there is no particular
need to impose a relation between $M_{\nu}$ and $M_{\Phi}$.

Representative
one-loop annihilation diagrams are shown in Fig.~\ref{fig:extras}. 
For a vector $s$-channel mediator, the $\gamma \gamma $ annihilation channel
is non-zero only if DM has axial couplings (in which case $\Phi$-mediated
diagrams also contribute; see the appendices for details).
This is one place in which our extended framework captures physics beyond what was
studied in Ref.~\cite{Jackson:2009kg}, where the DM was assumed to have 
vector-like couplings, and the $\gamma \gamma$ channel vanished.   In addition,
there are potentially gamma-ray line signals at various energies from 
annihilation into $\gamma Z$, $\gamma Z'$ and/or $\gamma \Phi$.
The $V \Phi$ final states proceed via $s$-channel vector exchange. 

\begin{table}[t]
\begin{center}
 $g_{\nu Z'}=g^V_{\nu Z'}$   \; \; \; \; \; \; \; \; \; \; \; \; \; \; \; \; \; \; \; \;  \; \; \; $g_{\nu Z'}=g^A_{\nu Z'}$ \;  \;  \;  \;  \;  \;  \; \; \; $g_{\nu Z'}=g^V_{\nu Z'}+g^A_{\nu Z'}$   \\ 
\begin{tabular}{|c|c|c|c|c|c|c|||c|c|c|c|||c|c|c|}
\hline
&&&&&&&&&&&&&\\
$g_{\psi Z}$ & V &V& V& A & V+A&V+A&V&V&V&V+A&V&V&V\\
\hline 
$g_{\psi Z'}$ & V & A&V+ A&V & V&V+A&V&A&V+A&V+A&V&A&V+A\\
\hline 
$\gamma \gamma$ & 0 &0& 0&0&0&0& 0&\Checkmark& \Checkmark&  \Checkmark&0&\Checkmark&\Checkmark\\
\hline
$\gamma Z$ & 0& \Checkmark& \Checkmark&  \Checkmark  & \Checkmark&  \Checkmark &0&\Checkmark& \Checkmark&  \Checkmark&0&\Checkmark&\Checkmark\\
\hline
$\gamma Z'$  & 0&0&\Checkmark& 0 & 0&  \Checkmark &0&0& \Checkmark&  \Checkmark &0 & 0&\Checkmark\\
\hline
$\gamma \Phi$  & 0 &0&\Checkmark&0 & 0&  \Checkmark &0&0& 0& 0&0&0&\Checkmark\\
\hline
$ZZ$  & 0& \Checkmark&\Checkmark&0& \Checkmark&  \Checkmark &0& \Checkmark& \Checkmark&  \Checkmark&0&\Checkmark&\Checkmark\\
\hline
$ZZ'$  & 0&0& \Checkmark & \Checkmark& \Checkmark&  \Checkmark & 0&0&\Checkmark&  \Checkmark&0&0&\Checkmark\\
\hline
$Z \Phi$  &0& \Checkmark&\Checkmark&0 &0&  \Checkmark &0& \Checkmark& \Checkmark&  \Checkmark &0&\Checkmark&\Checkmark\\
\hline
$Z' \Phi$  & 0& \Checkmark&\Checkmark&0 &0&  \Checkmark &0&0& \Checkmark&  \Checkmark &0&\Checkmark&\Checkmark\\
\hline
$\Phi \Phi$  & 0&0&0&0 & 0& 0&0&\Checkmark& \Checkmark&  \Checkmark&0&\Checkmark&\Checkmark\\
\hline
$Z'Z'$  & 0& \Checkmark&\Checkmark&0 & 0&  \Checkmark&0&\Checkmark& \Checkmark&  \Checkmark&0&\Checkmark&\Checkmark\\
\hline
\end{tabular}
\end{center}
\caption{\small One-loop DM annihilation channels in $U(1)^\prime$ model due to a single fermion $\psi$ running in the loop, where $\psi$
carries hypercharge 1 but is $ SU(2)$ singlet, for different combinations of vector (V) and axial (A) couplings.  Unlisted combinations lead to vanishing cross sections
in the non-relativistic limit.} 
\label{tab:unmixedZp}
\end{table}

We classify different scenarios based on whether interactions of $\nu$ and $\psi$
with the $Z^\prime$ are vector-like (V), axial-vector (A), or both (V+A).  The
various annihilation channels are listed along with their velocity-dependence, in
Table \ref{tab:unmixedZp}.
To recap the generic features:
\begin{itemize}
\item One-loop annihilations mediated by an $s$-channel scalar require pseudoscalar couplings of the DM.
\item One-loop annihilations into $\gamma \gamma$ mediated by an $s$-channel vector are possible 
only for axial couplings of both the DM and $\psi$.
\item One-loop annihilations into $\Phi \Phi$  require $s$-channel scalar exchange  with pseudoscalar coupling of DM and scalar coupling of $\psi$. 
\item The $\gamma V$ channel mediated by an $s$-channel scalar requires vector coupling of $\psi$ to $V$. 
\item The $\gamma \Phi$ channel requires vector couplings of both 
$\nu$ and $\psi$ to the $Z^\prime$.
\end{itemize}
The table displays the one-loop annihilation into 
$Z^\prime Z^\prime$ due to the $\psi$ loops. However, while we have explicitly
constructed a theory that is free of SM or mixed $U(1)^\prime$-SM anomalies,
we have not been careful to impose cancellation of the $U(1)^{\prime 3}$
anomaly.  Cancellation of such anomalies will generically require the addition of
fermions charged under $U(1)^\prime$ (but not the SM), and thus the rate into
$Z^\prime Z^\prime$ is somewhat more sensitive to the details of the UV completion.
Besides, this one-loop channel, as well as $Z'\Phi$ and $\Phi \Phi$, requires additional box diagrams that are not considered in this work. We will always work in the kinematical regime where this channel is absent.

\subsection{Vector DM Couplings}

\begin{figure}[t!]
\begin{center}
\includegraphics[angle=0,width=0.95\linewidth]{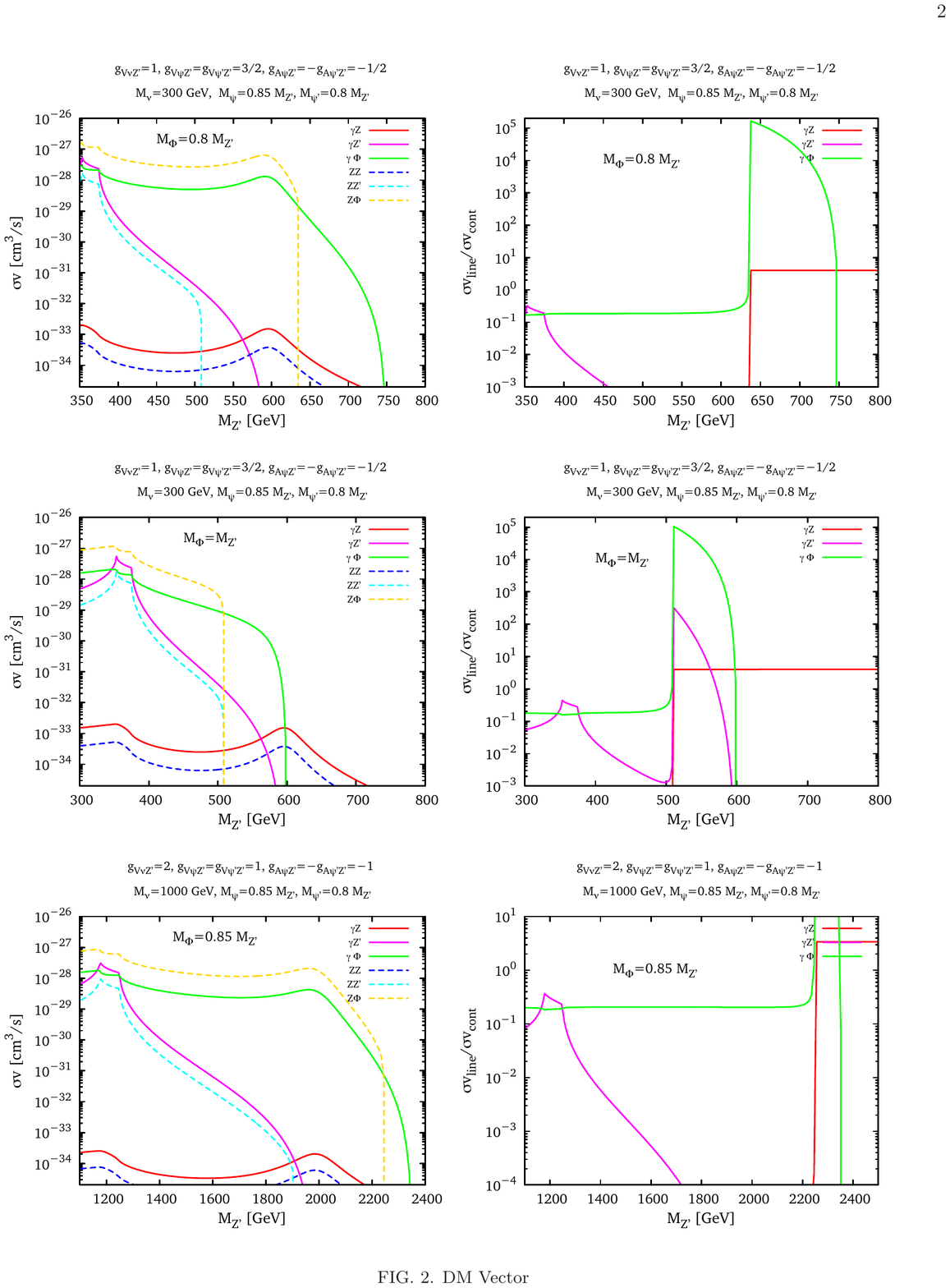}
\end{center}
\caption{ \small One-loop annihilation cross sections 
and ratio to continuum for DM with vector couplings, for a variety of parameter choices
as indicated. Ê}
\label{fig:VpA_VpA_Vector} 
\end{figure}

When the dark matter has vector-like coupling to the $Z^\prime$, a strong
line signal requires that $\psi$ have axial-vector, or a combination of vector and
axial vector couplings.  In this case, there are three distinct lines:
$\gamma Z$, $\gamma Z^\prime$, and $\gamma \Phi$.
We illustrate this case in Fig.~\ref{fig:VpA_VpA_Vector} for different 
configurations of parameters. 
Typically, the $\gamma Z^\prime$ and 
$\gamma \Phi$ cross sections are much larger than $\gamma Z$, with 
the $\gamma \Phi$ line very large for a wide range of $Z^\prime$ masses.

\begin{figure}[t!]
\begin{center}
\includegraphics[angle=0,width=0.995\linewidth]{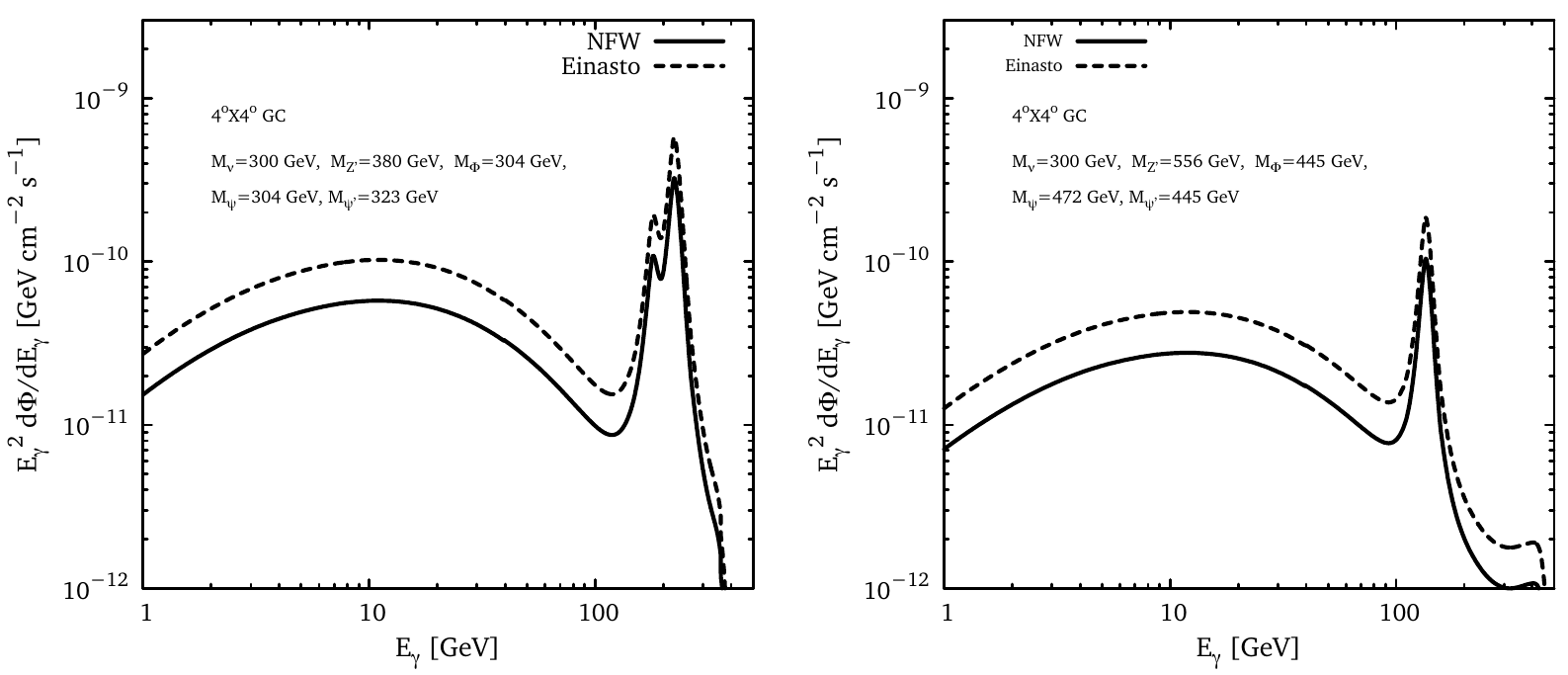}
\end{center}
\caption{ \small Gamma-ray spectrum obtained from $\nu \bar{\nu}$ annihilation in the galactic center  
with $M_{\nu}=300$~GeV and other parameters as indicated.}
\label{fig:spectrumDMvector} 
\end{figure}

We show two representative gamma ray spectra in 
Fig.~\ref{fig:spectrumDMvector} obtained for $M_{\nu}=300$ GeV.
In the first plot, there are two close-by lines, one at  an energy of $E_{\gamma}=180$ GeV  from 
$\gamma Z'$ 
($\sigma_{\gamma Z'} \ v= 4.7 \times 10^{-29}$ cm$^3$~s$^{-1}$), 
and one from  $\gamma \Phi$  at  $E_{\gamma}=223$ GeV with 
with $\sigma_{\gamma \Phi} \ v= 1.5 \times 10^{-28}$ cm$^3$~s$^{-1}$. 
The continuum is rather suppressed and dominated by $Z\Phi$ 
annihilation with $\sigma v= 8 \times 10^{-28}$ cm$^3$~s$^{-1}$.
In the second plot, the mass of $Z'$ is too large for the $\gamma Z'$ line to be observable 
($\sigma_{\gamma Z'} \ v=1.1 \times 10^{-33} $ cm$^3$~s$^{-1}$) but there is a striking line at  $E_{\gamma}=135$ GeV and with $\sigma_{\gamma \Phi} \ v= 7.4 \times 10^{-29}$ cm$^3$~s$^{-1}$. The continuum dominated by the $Z \Phi$ channel has  $\sigma_{Z \Phi} \ v= 3.9 \times 10^{-28}$ cm$^3$~s$^{-1}$. 

\subsection{Axial-Vector DM Couplings}

\begin{figure}[t!]
\begin{center}
\includegraphics[angle=0,width=0.8\linewidth]{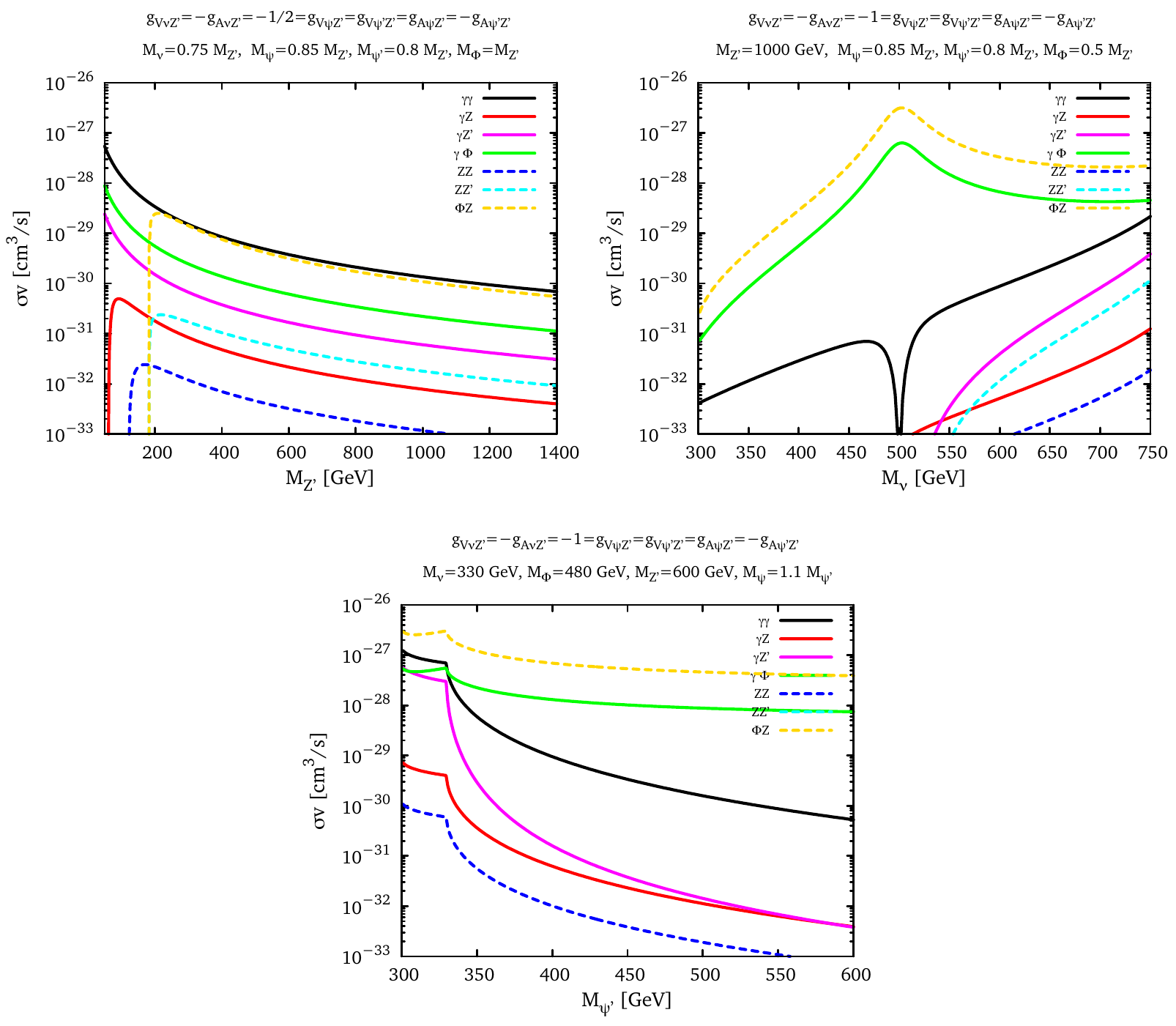}
\end{center}
\caption{ \small  Cross sections for models where the DM has axial vector couplings,
as a function of: a) $\langle \Phi \rangle$ (with all couplings held fixed); 
b) $y^S_{\nu \Phi}$/$M_\nu$ (with all other parameters held fixed);
c) $y^S_{\psi \Phi}$/$M_\psi$  (with all other parameters held fixed). }
\label{fig:DMaxial} 
\end{figure}

When the DM has $Z^\prime$ axial vector couplings, it opens the door for
the $\gamma \gamma$ channel and in addition there are contributions from
$\Phi$ exchange to the annihilation channels.
We illustrate this case in Figures~~\ref{fig:DMaxial}, where we have chosen to switch off the pseudo-scalar
coupling of the DM, and thus
all diagrams mediated by $s$-channel scalar exchange vanish. 

In Figure~\ref{fig:DMaxial}, we show the dependence of the cross sections
on $M_{Z^\prime}$, $M_\nu$, and $M_{\psi}$.  Because all three of these
parameters are proportional to $\langle \Phi \rangle$, we vary them in a
controlled way, exploring successively the variation with changing 
$\langle \Phi \rangle$ and all couplings fixed, variation of $M_\nu$ obtained
by varying $y^S_{\nu \Phi}$ with all other parameters fixed, and variation of
$M_\psi$ obtained by varying $y^S_{\psi \Phi}$ with all other parameters fixed.
From the figures, we see that either the $\gamma \gamma$ or $\gamma \Phi$
may end up as the strongest of the line signals (with $\gamma \gamma$ 
vanishing when the $Z^\prime$ goes on-shell, as required by the Landau-Yang
theorem \cite{Yang:1950rg}), whereas the one loop
continuum is rather anemic and is
almost entirely made up of the $Z\Phi$ channel.

\begin{figure}[b!]
\begin{center}
\includegraphics[angle=0,width=0.4\linewidth]{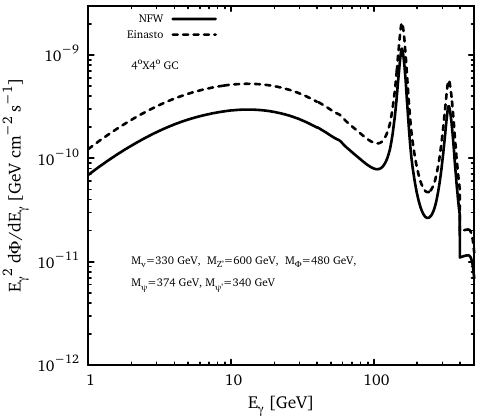}
\end{center}
\caption{ \small Two-peak Gamma-ray spectrum obtained from $\nu \bar{\nu}$ annihilation in the galactic center with $M_{\nu}=330$~GeV and other
parameters as shown. }
\label{fig:spectrumDMaxial} 
\end{figure}

We show a representative two-peak gamma ray spectrum in 
Figure~\ref{fig:spectrumDMaxial} for a case with $M_{\nu}=330$~GeV.
For these parameters, the $\gamma Z^\prime$ line is somewhat 
too faint to be observed compared to the $Z \Phi$ continuum 
($\sigma_{\gamma Z'} \ v= 1 \times 10^{-29}$ cm$^3$~s$^{-1}$ 
and $\sigma_{Z \Phi} \ v= 4.5 \times 10^{-27}$ cm$^3$~s$^{-1}$), 
but there are two large lines from the
$\gamma \Phi$ and $\gamma \gamma$ channels at 
energies of $E_{\gamma}= 155$ and $330$ GeV, respectively, with cross sections
$\sigma_{\gamma \Phi} \ v= 8.5 \times 10^{-28}$ cm$^3$~s$^{-1}$ and
$\sigma_{\gamma \gamma} \ v= 1.2 \times 10^{-28}$ cm$^3$~s$^{-1}$.
%
%
%

\subsection{A Purely Vector-Like Model} 

\begin{figure}[t!]
\begin{center}
\includegraphics[angle=0,width=0.55\linewidth]{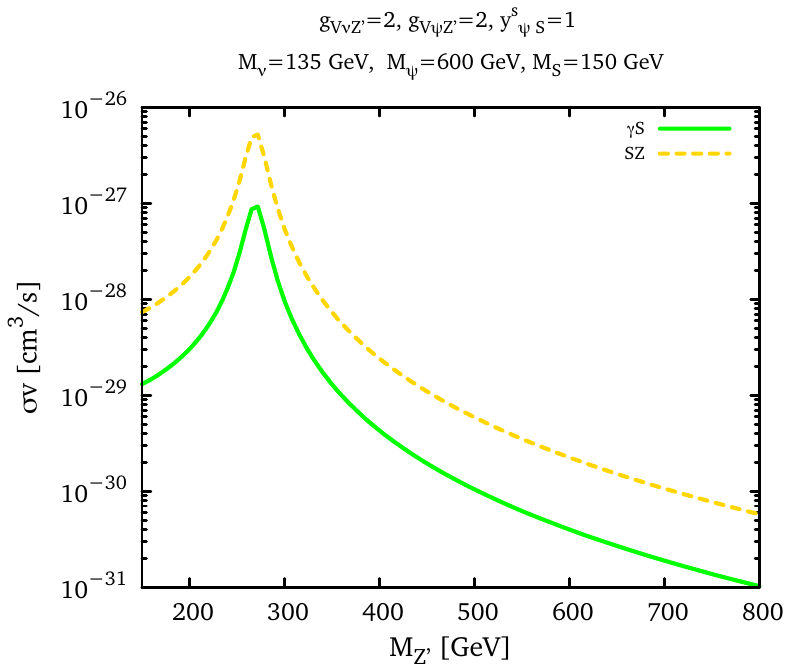}
\end{center}
\caption{ \small One-loop annihilation cross sections into $\gamma S$ and $Z S$, for the purely vector-like
$Z^\prime$ model.}
\label{fig:allvector} 
\end{figure}

There is one further ingredient that is worthwhile to consider adding on top of the
minimal framework we have introduced so far.  In theories where the $\nu$ and
$\psi$ fields are composites, the most natural situation is one with 
only vector-like couplings.  As shown in Table~\ref{tab:unmixedZp}, this is not
expected to lead to any observable line signals.  However, such
models often also predict additional scalars which are unrelated to the 
$U(1)^\prime$ breaking, and are singlets under the SM, but 
have scalar interactions with the
$\psi$ fields.  For example, composite Higgs
models based on the $SO(6) \to SO(5)$ symmetry-breaking contain
such a state $S$ \cite{Gripaios:2009pe,Espinosa:2011eu}.  
Such an additional state opens up the possibility of strong gamma ray lines 
from the $\gamma S$ channel mediate by $Z^\prime$ exchange.
In addition, one would also expect annihilation into $Z S$ and 
$Z^\prime S$.  In order to simplify the discussion,
we assume that the DM coupling to $S$ is scalar, and thus any additional mediation
via $S$ itself vanishes in the non-relativistic limit.  
This case is summarized in Table~\ref{tab:allvector} and 
illustrated with some representative quantitative results in Figure~\ref{fig:allvector}.
\begin{table}[!h]
\begin{center}
$\lambda_{\nu}$: scalar  \; \; \; \; \; \; \; \;  \; \; \;$\lambda_{\nu}$: pseudoscalar\\
\begin{tabular}{|c|c|c|||c|c|}
\hline
$\lambda_{\psi}$ & scalar & pseudoscalar &  scalar & pseudoscalar\\
\hline 
$\gamma \gamma$ &0&0&\Checkmark&\Checkmark\\ 
\hline 
$\gamma Z $&0&0&\Checkmark&\Checkmark\\
\hline 
$\gamma Z'$ &0&0&\Checkmark&\Checkmark\\
\hline 
$\gamma S$  & \Checkmark &\Checkmark & \Checkmark& \Checkmark \\
\hline
$ZZ$ &0&0&\Checkmark&\Checkmark \\
\hline 
$ZZ'$ &0&0&\Checkmark&\Checkmark \\
\hline 
$Z S$  &\Checkmark&0 &\Checkmark &0 \\
\hline
$Z' S$  & \Checkmark& 0 & \Checkmark&0 \\
\hline 
$SS$  & \Checkmark& 0 & \Checkmark&0 \\
\hline 
$Z'Z'$  & 0& 0 & \Checkmark&\Checkmark \\
\hline
\end{tabular}
\end{center}
\caption{ \small One-loop DM annihilation channels in the purely vector-like $U(1)^\prime$ model with an
additional gauge singlet scalar particle $S$. } 
\label{tab:allvector}
\end{table}

\subsection{Summary}

To summarize this section, we have seen that in a simple model with vector mediator, there is a vast range of possibilities for large line signals depending on the nature of the couplings and mass relations. 
A key-result is that axial couplings are
required to have non-vanishing 1-loop cross sections into two gauge bosons.
Gamma ray spectra with several lines are typical in the presence of axial couplings.
We have presented a minimal anomaly-free construction that realizes non-vanishing axial couplings. 
It relies on  two heavy fermions $\psi$ and $\psi'$ with equal vector couplings and opposite axial couplings to $Z'$.

If $M_{\psi}=M_{\psi'}$, the annihilation channels into two gauge bosons, namely into $\gamma Z$, $\gamma Z'$, $ZZ$ and $ZZ'$, vanish as the two diagrams with respectively $\psi$ and $\psi'$ in the loop cancel exactly.
On the other hand,  for $\gamma \Phi$ and $Z \Phi$ channels,  no such cancellation occurs as these channels do not require axial couplings.
In general, $M_{\psi}$ and $M_{\psi'}$ are different, however there is some partial remnant cancellation which reduces the $\gamma Z$ and $ZZ$ channels relatively to the others.
The $\gamma Z'$ and $ZZ'$ channels are not as suppressed 
(when kinematically allowed) because of the larger couplings involved,
$ g_{\psi,Z'}\sim 10 g_{\psi,Z}$, and two possible axial vertices.

We therefore conclude that quite generally, the  $Z \Phi$ and $\gamma \Phi$ channels are the dominant 1-loop annihilation processes.

Similarly to the scalar mediator case, the relic density is complicated and one
could imagine several different mechanisms which could arrange for a thermal relic while simultaneously resulting
in large gamma ray line signals, but there are no firm generic correlations between the two.  Some discussion of the
relic density in vector-mediated models can be found in \cite{Jackson:2013tca}.

\section{Conclusion}
\label{sec:summary}

Very large gamma ray line signals would motivate
models where DM annihilation rates are enhanced by resonant effects and where 
the continuum is depressed because of annihilation into heavy channels.  
In this work we have explored the two main categories which fit easily into
the framework of a renormalizable quantum field theory:
vector-resonance and scalar-resonance models.
We have introduced simple effective field theory sketches of such theories
where the dark matter is a Dirac fermion, and
performed the first complete analysis which includes the one loop contributions to
the continuum annihilation modes.  

Both cases, with appropriate choices of couplings, can lead to large line signals 
as well as large line/continuum ratios.
For $s$-channel  vector mediators, this requires 
chiral couplings to the fermions running in the loop, whereas for
scalar mediators, pseudo-scalar couplings with DM are necessary.  The various cases
are summarized in tables~\ref{tab:unmixedPhi} and \ref{tab:unmixedZp}.
We have restricted our models to those in which the fermions running in the
loops are disconnected from the SM fermions.  We explore the case
in which the loop fermions mix with the top quark in a separate
publication \cite{Jackson:2013tca}.

Gamma ray lines are a fascinating, powerful probe of dark matter annihilations.
Mapping out the kinds of theories which naturally produce such features
is an important step in preparing for a discovery.

\section*{Acknowledgments}
G. Servant is supported by the ERC starting grant Cosmo@LHC (204072).
She thanks Sean Tulin for discussions.
G. Shaughnessy is supported by the U. S. Department of Energy under the contract 
DE-FG-02-95ER40896.  He also thanks W.-Y. Keung for discussions.
T. Tait acknowledges the
hospitality of the SLAC theory group, and is supported in part by NSF
grant PHY-0970171. He also thanks P. Fox, J. Kearny, and A. Pierce for early 
collaboration and discussions, and
the Aspen Center for Physics, under NSF Grant No. 1066293, where part of this work 
was completed. 
The work of M. Taoso is supported by the European Research Council (ERC)
under the EU Seventh Framework Programme (FP7/2007-2013) / ERC
Starting Grant (agreement n. 278234 - `NewDark' project).


\appendix

\section{Effective Vertices and Amplitudes-squared}
\label{app:effective-vertices}

\begin{figure}[t]
\begin{center}
\includegraphics[scale=0.75,bb = 115 454 493 715]{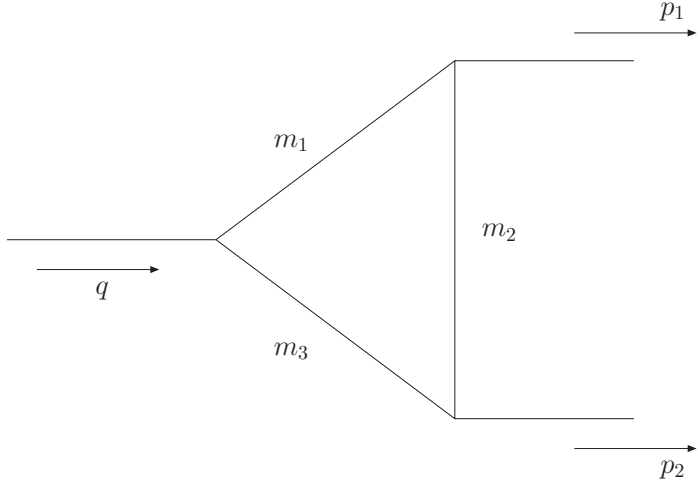} 
\caption[]{Topology for the effective vertices.}
\label{fig:1-to-2}
\end{center}
\end{figure}

In the following appendices, we summarize the one-loop expressions for the effective vertices and amplitudes-squared needed for the calculations performed here.  The generic topology for the loop diagrams considered here is shown in Fig.~\ref{fig:1-to-2}.  We express all amplitudes in terms of two-point ($B_0$) and three-point ($C_0$) scalar integrals where:
\begin{eqnarray}
C_0 &=& C_0(M_1^2, M_2^2, 4 M_\nu^2; m_1^2, m_2^2, m_3^2) \, , \\
B_0(23) &=& B_0(M_2^2; m_2^2, m_3^2) \, ,\\
B_0(13) &=& B_0(4 M_\nu^2; m_1^2, m_3^2) \, ,\\
B_0(12) &=& B_0(M_1^2; m_1^2, m_2^2) \, .
\end{eqnarray}
In the following expressions, for the sake of simplicity, we have set all loop masses equal ($m_1 = m_2 = m_3 \equiv m_f$).  However, to derive our results, we have computed all of the expressions in these appendices in terms of the general masses as depicted in Fig.~\ref{fig:1-to-2}.

%
%
\section{One-loop annihilation into $\gamma \gamma$ (and $gg$)}
\label{app:gamma-gamma}

The annihilation channel $\nu \bar{\nu} \to \gamma \gamma$ can proceed via $s$-channel exchange of both a $Z^\prime$ as well as the scalar $\Phi$.  The effective vertex for $\Phi(q) \to \gamma^\mu(p_1) \gamma^\nu(p_2)$ takes the form:
\begin{equation}
{\cal{V}}_{\Phi \to \gamma\gamma}^{\mu\nu} = A g^{\mu\nu}  + B p_1^\nu p_2^\mu + i C \epsilon^{p_1 p_2 \nu \mu} \,,
\end{equation}
where the coefficients $A, B$ and $C$ are:
\begin{eqnarray}
A &=& -\frac{v_{(1,2)}^\gamma v_{(2,3)}^\gamma y_{s,(1,3)}^\Phi m_f \left(2 C_0
   \left(m_f^2-M_\nu^2\right)+1\right)}{2 \pi ^2} \\
\nonumber\\
B &=& \frac{v_{(1,2)}^\gamma v_{(2,3)}^\gamma y_{s,(1,3)}^\Phi m_f \left(2 C_0
   \left(m_f^2-M_\nu^2\right)+1\right)}{4 \pi ^2 M_\nu^2} \\
\nonumber\\
C &=& \frac{y_{p, (1,3)}^\Phi v_{(1,2)}^\gamma v_{(2,3)}^\gamma m_f C_0}{2 \pi ^2}
\end{eqnarray}
The couplings $y_{s,f\bar{f}}^\Phi$ and $y_{p,f\bar{f}}^\Phi$ are the scalar and pseudoscalar couplings, respectively, of $\Phi$ to the fermions running in the loop, while $v_{(i,j)}^V$ ($a_{(i,j)}^V$) are the vector (axial-vector) couplings between fermion pair $(i,j)$ and gauge boson $V$.

The effective vertex for $Z^\prime \to \gamma \gamma$ can be written as:
\begin{equation}
{\cal{V}}_{Z^\prime \to \gamma \gamma}^{\alpha \mu \nu} = D \epsilon^{p_1 \mu \nu \alpha} \,,
\end{equation}
where we have used transversality of the photons (i.e., $\epsilon(p_1) \cdot p_1 = \epsilon(p_2) \cdot p_2 = 0$) as well as the fact that the photons are identical particles to eliminate other possible tensor structures.  The loop coefficient $D$ is given by:
\begin{eqnarray}
D = \frac{a_{f\bar{f}}^{Z^\prime}}{4 \pi^2} \left[ 1 + 2 m_f^2 C_0 \right] \,.
\end{eqnarray}  

Averaging over initial spins, summing over final spins and including a factor of 1/2 for identical particles in the final state, the total matrix-element-squared for $\nu + \bar{\nu} \to \gamma \gamma$ is given by:
\begin{equation}
\overline{\sum} \left| {\cal M}_{\nu \bar{\nu} \to \Phi \Phi} \right|^2 = \left| {\cal M}_\Phi \right|^2 + \left| {\cal M}_{Z^\prime} \right|^2 + 2 Re \left| M_\Phi \cdot M_{Z^\prime}^* \right| \,,
\end{equation}
where the matrix-element-squared for the $\Phi$-mediated amplitude is:
\begin{eqnarray}
\overline{\sum} |{\cal M}|_\Phi^2 &=& \frac{16 \left(y_{p,\nu\bar{\nu}}^\Phi\right)^2 M_\nu^2}{\left| \Sigma_\Phi \right|^2} \left[ | A |^2 + 4 M_\nu^4 | C |^2 \right] \,.
\end{eqnarray} 
The denominator factor $\Sigma_\Phi$ is defined as:
\begin{equation}
\Sigma_\Phi = 4 M_\nu^2 - M_\Phi^2 + i M_\Phi \Gamma_\Phi \, .
\label{eq:Sigma_Phi}
\end{equation}
The amplitude-squared for the $Z^\prime$-mediated channel is:
\begin{eqnarray}
\overline{\sum} |{\cal M}|_{Z^\prime}^2 &=& \frac{64 \left( a_{\nu\bar{\nu}}^{Z^\prime} \right)^2 M_\nu^4 \left(4 M_\nu^2 - M_{Z^\prime}^2\right)^2}{M_{Z^\prime}^4 \left| \Sigma_{Z^\prime} \right|^2} \, | D |^2 \,,
\end{eqnarray}
where:
\begin{equation}
\Sigma_{Z^\prime} = 4 M_\nu^2 - M_{Z^\prime}^2 + i M_{Z^\prime} \Gamma_{Z^\prime} \, .
\label{eq:Sigma_Zp}
\end{equation}
It is interesting to note that, in the limit where the $Z^\prime$ goes on-shell, we replace $4 M_\nu^2 \to M_{Z^\prime}^2$ and the amplitude-squared vanishes as required by the Landau-Yang theorem. 
The cross term between the $\Phi$- and $Z^\prime$-mediated processes is given by:
\begin{eqnarray}
2 Re \left| M_\Phi \cdot M_{Z^\prime}^* \right| &=& - Re \biggl[ \frac{64 y_{p,\nu\bar{\nu}}^\Phi a_{\nu\bar{\nu}}^{Z^\prime}  M_\nu^5 \left(4 M_\nu^2 - M_{Z^\prime}^2\right)}{M_{Z^\prime}^2 \Sigma_\Phi \Sigma_{Z^\prime}^* } \, \left( C D^* + C^* D \right) \biggr] \,,
\end{eqnarray}

Finally, we note that we can obtain the $\nu \bar{\nu} \to gg$ channel with the simple rescaling:
\begin{eqnarray}
\sigma_{gg} = \left( \frac{2}{9} \right) \left( \frac{\alpha}{\alpha_s} \right)^2 \frac{1}{Q_t^2} \, \sigma_{\gamma\gamma} \,,
\end{eqnarray}
where $Q_t$ is the charge of the top quark in units of $e$ (i.e., $Q_t = 4/3$).

%
%
\section{One-loop annihilation into $\Phi \Phi$}
\label{app:phi-phi}

The annihilation channel $\nu \bar{\nu} \to \Phi \Phi$ can proceed via both $s$-channel $\Phi$ and $Z^\prime$ exchange.  The effective vertex for the $\Phi \Phi \Phi$ vertex takes the form:
\begin{equation}
{\cal{V}}_{\Phi \to \Phi\Phi} = A
\end{equation}
where:
\begin{eqnarray}
A &=& \frac{i m_f}{2\pi^2} \biggl[
y_{s,f\bar{f}}^\Phi \left(y_{p,f\bar{f}}^\Phi \right)^2
\biggl( B_0(13) + 2 B_0(23) - (2 M_\nu^2 + M_\Phi^2) C_0 \biggr) \nonumber\\
&& \,\,\,\,\,\,\,\,\, -
\left(y_{s,f\bar{f}}^\Phi \right)^3
\biggl( B_0(13) + 2 B_0(23) - (2 M_\nu^2 - 4 m_f^2 + M_\Phi^2) C_0 \biggr)
\biggr]
\end{eqnarray}
while the $Z^\prime \Phi \Phi$ effective vertex is given by:
\begin{equation}
{\cal{V}}^\alpha_{Z^\prime \to \Phi\Phi} = B p_1^\alpha + C p_2^\alpha \,.
\end{equation}
The loop coefficients ($B$ and $C$) in this case are:
\begin{equation}
B = C =  -\frac{i a_{f\bar{f}}^{Z^\prime} y_{p,f\bar{f}}^\Phi  y_{s,f\bar{f}}^\Phi m_f^2 C_0 }{\pi ^2} \,.
\end{equation}

The amplitude-squared averaged/summed over initial/final spins takes the form:
\begin{equation}
\overline{\sum} \left| {\cal M}_{\nu \bar{\nu} \to \Phi \Phi} \right|^2 = \left| {\cal M}_\Phi \right|^2 + \left| {\cal M}_{Z^\prime} \right|^2 + 2 Re \left| M_\Phi \cdot M_{Z^\prime}^* \right| \,,
\end{equation}
where the individual parts are:
\begin{eqnarray}
\left| {\cal M}_\Phi \right|^2 &=&  \frac{\left(y_{p,\nu\bar{\nu}}^\Phi \right)^2 M_\nu^2}{\left|\Sigma_\Phi \right|^2} \left| A \right|^2 \,, \\
\nonumber\\
\left| {\cal M}_{Z^\prime} \right|^2 &=& \frac{ 1}{M_{Z^\prime}^4 \left|\Sigma_{Z^\prime} \right|^2} M_\nu^2 M_{Z^\prime}^4 \biggl(M_\nu^2 \biggl(\left(a_{\nu\bar{\nu}}^{Z^\prime} \right)^2
   (B+C)
   (B^*+C^*) \nonumber\\
&& + \left(v_{\nu\bar{\nu}}^{Z^\prime} \right)^2
   (B-C)
   (B^*-C^*)\biggr)-\left(v_{\nu\bar{\nu}}^{Z^\prime} \right)^2 M_\Phi^2
   (B-C)
   (B^*-C^*)\biggr) \nonumber\\
&&+16 \left(a_{\nu\bar{\nu}}^{Z^\prime} \right)^2 M^8
   (B+C) (B^*+C^*) \nonumber\\
&&-8
   \left(a_{\nu\bar{\nu}}^{Z^\prime} \right)^2 M^6 M_{Z^\prime}^2 (B+C)
   (B^*+C^*) \,, \\
\nonumber\\
2 Re \left| M_\Phi \cdot M_{Z^\prime}^* \right| &=& Re \biggl[ \frac{ 2 y_{p,\nu\bar{\nu}}^\Phi v_{\nu\bar{\nu}}^{Z^\prime} M_\nu^3 \left( 4 M_\nu^2 - M_{Z^\prime}^2 \right)}{M_{Z^\prime}^2 \Sigma_\Phi \Sigma_{Z^\prime}^*} A \left( B^* + C^* \right) \biggr]
\end{eqnarray}
Note that a consistent 1-loop calculation of the $\nu \bar{\nu} \to \Phi \Phi $ channel requires the addition of box diagrams.
This is beyond the scope of this paper which is focussed on the topology shown in Fig.~\ref{fig:1-to-2}. In all our illustrative plots, we assume $M_{\nu}< M_{\phi}$.

%
%
\section{ One-loop annihilation into $V \gamma$}
\label{app:V-gamma}

The annihilation $\nu \bar{\nu} \to V \gamma$ (where $V$ = $Z$ or $Z^\prime$) can proceed via both $\Phi$ and $Z^\prime$ $s$-channel diagrams.  The effective vertices take the forms:
\begin{equation}
{\cal{V}}_{\Phi \to V \gamma}^{\mu \nu} = A g^{\mu \nu} + B p_2^\mu p_1^\nu + i C \epsilon^{p_1 p_2 \mu \nu} \, , 
\end{equation}
and:
\begin{equation}
{\cal{V}}_{Z^\prime \to V \gamma}^{\alpha \mu \nu} = D \epsilon^{p_1 \mu \nu \alpha} + E \epsilon^{p_2 \mu \nu \alpha} + F p_1^\alpha \epsilon^{p_1 p_2 \mu \nu} + G p_2^\alpha \epsilon^{p_1 p_2 \mu \nu} + H p_1^\nu \epsilon^{p_1 p_2 \alpha \mu} + I p_2^\mu \epsilon^{p_1 p_2 \alpha \nu} \,.
\end{equation}

The loop coefficients are:
\begin{eqnarray}
A &=& \frac{v_{f\bar{f}}^\gamma v_{f\bar{f}}^V y_{s, f\bar{f}}^\Phi m_f }{4 \pi ^2
   \biggl(4 M_\nu^2-M_V^2\biggr)}\biggl(M_V^2
   \biggl(-2 B_0(13)+2 B_0(23)+C_0 M_V^2+4 C_0
   m_f^2+2\biggr) \nonumber\\
&&\,\,\,\,\,\,\,\,\,\,\,\,\,\,\,\,\,\,\,\,\,\,+16 C_0 M_\nu^4-8 M_\nu^2 \biggl(C_0
   \biggl(M_V^2+2 m_f^2\biggr)+1\biggr)\biggr) \,, \\
\nonumber\\
B &=& \frac{v_{f\bar{f}}^\gamma v_{f\bar{f}}^V y_{s, f\bar{f}}^\Phi m_f}{2 \pi ^2
   \biggl(M_V^2-4 M_\nu^2\biggr)^2} \biggl(-M_V^2
   \biggl(-2 B_0(13)+2 B_0(23)+C_0 M_V^2+4 C_0
   m_f^2+2\biggr) \nonumber\\
&&\,\,\,\,\,\,\,\,\,\,\,\,\,\,\,\,\,\,\,\,\,\,-16 C_0 M_\nu^4+8 M_\nu^2 \biggl(C_0
   \biggl(M_V^2+2 m_f^2\biggr)+1\biggr)\biggr) \, , \\
\nonumber \\
C &=& \frac{y_{p,f\bar{f}}^\Phi v_{f\bar{f}}^\gamma v_{f\bar{f}}^{V} m_f C_0}{2 \pi ^2} \,, \\
\nonumber\\
D &=& \frac{v_{f\bar{f}}^\gamma}{8 \pi ^2} \biggl(\frac{2 B_0(12) M_V^2 (a_{f\bar{f}}^V
   v_{f\bar{f}}^{Z^\prime}+a_{f\bar{f}}^{Z^\prime} v_{f\bar{f}}^V)}{M_V^2-4
   M_\nu^2}+\frac{B_0(13) \biggl(4 M_\nu^2+M_V^2\biggr)^2 (a_{f\bar{f}}^V
   v_{f\bar{f}}^{Z^\prime}+a_{f\bar{f}}^{Z^\prime} v_{f\bar{f}}^V)}{\biggl(M_V^2-4
   M_\nu^2\biggr)^2} \nonumber\\
&&-\frac{2 B_0(23) M_V^2 \biggl(4
   M_\nu^2+M_V^2\biggr) (a_{f\bar{f}}^V v_{f\bar{f}}^{Z^\prime}+a_{f\bar{f}}^{Z^\prime}
   v_{f\bar{f}}^V)}{\biggl(M_V^2-4 M_\nu^2\biggr)^2}+2 (B_1(12)+1)
   (a_{f\bar{f}}^V v_{f\bar{f}}^{Z^\prime}+a_{f\bar{f}}^{Z^\prime} v_{f\bar{f}}^V) \nonumber\\
&&+4 C_0 m_f^2
   (a_{f\bar{f}}^{Z^\prime} v_{f\bar{f}}^V-a_{f\bar{f}}^V v_{f\bar{f}}^{Z^\prime})-\frac{8 C_0
   a_{f\bar{f}}^V v_{f\bar{f}}^{Z^\prime} M_\nu^2 M_V^2}{M_V^2-4 M_\nu^2}-\frac{8
   C_0 a_{f\bar{f}}^{Z^\prime} v_{f\bar{f}}^V M_\nu^2 M_V^2}{M_V^2-4
   M_\nu^2}\biggr) \, , \\
\nonumber\\
E &=&-\frac{v_{f\bar{f}}^\gamma (a_{f\bar{f}}^V v_{f\bar{f}}^{Z^\prime}+a_{f\bar{f}}^{Z^\prime} v_{f\bar{f}}^V)}{8 \pi ^2 \biggl(4 M_\nu^2-M_V^2\biggr)} \times \nonumber\\
&& \,\,\,\,\,\,\,\,\,\,\,\,\,\,\,\,\,\,\,\,\,\,  \biggl(M_V^2 \biggl(B_0(13)-B_0(23)-2 C_0
   m_f^2+1\biggr)+M_\nu^2 \biggl(8 C_0
   m_f^2-4\biggr)\biggr) \, ,\\
\nonumber\\
F &=& -\frac{v_{f\bar{f}}^\gamma (a_{f\bar{f}}^V v_{f\bar{f}}^{Z^\prime}+a_{f\bar{f}}^{Z^\prime} v_{f\bar{f}}^V)}{4 \pi ^2 \biggl(4
   M_\nu^2-M_V^2\biggr)^3}
   \biggl(B_0(13) \biggl(16 M_\nu^4+16 M_\nu^2 M_V^2+M_V^4\biggr) \nonumber\\
&&-2
   \biggl(M_V^2 \biggl(4 M_\nu^2 \biggl(B_0(12)+2
   B_0(23)+C_0 M_V^2 -2 C_0
   m_f^2-1\biggr) \nonumber\\
&&+M_V^2 \biggl(-B_0(12)+B_0(23)+2
   C_0 m_f^2+1\biggr)-16 C_0 M_\nu^4\biggr) \nonumber\\
&&-B_1(12)
   \biggl(M_V^2-4 M_\nu^2\biggr)^2\biggr)\biggr) \,, \\
\nonumber
\end{eqnarray}
\begin{eqnarray}
G &=& \frac{v_{f\bar{f}}^\gamma (a_{f\bar{f}}^V v_{f\bar{f}}^{Z^\prime}+a_{f\bar{f}}^{Z^\prime} v_{f\bar{f}}^V)}{4 \pi ^2 \biggl(M_V^2-4
   M_\nu^2\biggr)^2} \times \nonumber\\
&&\,\,\,\,\,\,\,\,\,\,\,\,\,\,\,\,\,\,\,\,\,\, 
   \biggl(M_V^2 \biggl(B_0(13)-B_0(23)-2 C_0
   m_f^2-1\biggr)+M_\nu^2 \biggl(8 C_0
   m_f^2+4\biggr)\biggr) \,, \\
\nonumber
\end{eqnarray}
\begin{eqnarray}
H &=& -\frac{v_{f\bar{f}}^\gamma (a_{f\bar{f}}^V v_{f\bar{f}}^{Z^\prime}+a_{f\bar{f}}^{Z^\prime} v_{f\bar{f}}^V)}{4 \pi ^2
   \biggl(4 M_\nu^2-M_V^2\biggr)^3} \biggl(-8
   M_\nu^2 M_V^2 \biggl(B_0(12)-3 B_0(13)+3 B_0(23)+2
   B_1(12) \nonumber\\
&&+C_0 M_V^2+2 C_0
   m_f^2+1\biggr)+M_V^4 (2 B_0(12)-B_0(13)+2
   B_1(12)) \nonumber\\
&&+16 M_\nu^4 \biggl(B_0(13)+2 \biggl(B_1(12)+C_0
   M_V^2+2 C_0 m_f^2+1\biggr)\biggr)\biggr) \,, \\
\nonumber\\
I &=& \frac{v_{f\bar{f}}^\gamma (a_{f\bar{f}}^V v_{f\bar{f}}^{Z^\prime}+a_{f\bar{f}}^{Z^\prime} v_{f\bar{f}}^V)}{4 \pi ^2 \biggl(M_V^2-4
   M_\nu^2\biggr)^2} \times \nonumber\\
&&  \,\,\,\,\,\,\,\,\,\,\,\,\,\,\,\,\,\,\,\,\,\,  \biggl(M_V^2 \biggl(B_0(13)-B_0(23)-2 C_0
   m_f^2-1\biggr)+M_\nu^2 \biggl(8 C_0
   m_f^2+4\biggr)\biggr)
\end{eqnarray}

The amplitude-squared averaged/summed over initial/final spins takes the form:
\begin{equation}
\overline{\sum} \left| {\cal M}_{\nu \bar{\nu} \to V \gamma} \right|^2 = \left| {\cal M}_\Phi \right|^2 + \left| {\cal M}_{Z^\prime} \right|^2 + 2 Re \left| M_\Phi \cdot M_{Z^\prime}^* \right| \,,
\end{equation}
where the individual parts are:
\begin{eqnarray}
\left| {\cal M}_\Phi \right|^2 &=& \frac{4 \left( y_{p,\nu\bar{\nu}}^\Phi \right)^2 M_\nu^2}{\left| \Sigma_\Phi \right|^2} \left(4 \left| A \right|^2+\left| C \right|^2 \left(M_V^2-4
   M_\nu^2\right)^2\right) \,,\\
\nonumber
\end{eqnarray}
\begin{eqnarray}
\left| {\cal M}_{Z^\prime} \right|^2 &=& \frac{1}{ \left| \Sigma_{Z^\prime} \right|^2 } \biggl[
  \left(v_{\nu\bar{\nu}}^{Z^\prime}\right)^2 \biggl(2 (D E^* + D^* E ) \left(4 M_\nu^2-M_V^2\right) \left(12
   M_\nu^2+M_V^2\right) \nonumber\\
&&+(D F^* + D^* F ) \left(4
   M_\nu^2-M_V^2\right)^3+(D G^* + D^* G ) \left(M_V^2-4
   M_\nu^2\right)^3 \nonumber\\
&&+\frac{4 (D H^* + D^* H ) M_\nu^2 \left(4
   M_\nu^2-M_V^2\right)^3}{M_V^2}+\frac{| D |^2 \left(4
   M_\nu^2+M_V^2\right) \left(M_V^2-4
   M_\nu^2\right)^2}{M_V^2} \nonumber\\
&&+(E F^* + E^* F ) \left(4 M_\nu^2+M_V^2\right)
   \left(M_V^2-4 M_\nu^2\right)^2 \nonumber\\
&&-(E G^* + E^* G ) \left(4
   M_\nu^2+M_V^2\right) \left(M_V^2-4 M_\nu^2\right)^2+8 (E H^* + E^* H )
   M_\nu^2 \left(M_V^2-4 M_\nu^2\right)^2 \nonumber\\
&&+| E |^2 \left(16 M_\nu^4+24 M_\nu^2
   M_V^2+M_V^4\right)+\frac{1}{2} (F G^* + F^* G ) \left(4
   M_\nu^2-M_V^2\right)^3 \left(M_V^2-4 M_\nu^2\right) \nonumber\\
&&+\frac{1}{4}
   | F |^2 \left(M_V^2-4 M_\nu^2\right)^4+\frac{1}{4} | G |^2
   \left(M_V^2-4 M_\nu^2\right)^4+\frac{| H |^2 M_\nu^2
   \left(M_V^2-4 M_\nu^2\right)^4}{M_V^2}\biggr) \nonumber\\
&+&
\left(a_{\nu\bar{\nu}}^{Z^\prime}\right)^2
   \biggl(-\frac{2 (D E^* + D^* E ) \left(4 M_\nu^2-M_V^2\right)^2
   \left(M_{Z^\prime}^2-4 M_\nu^2\right)^2}{M_{Z^\prime}^4} \nonumber\\
&&+\frac{(D F^* + D^* F )
   \left(4 M_\nu^2-M_V^2\right)^3 \left(M_{Z^\prime}^2-4
   M_\nu^2\right)^2}{M_{Z^\prime}^4} \nonumber\\
&&+\frac{(D G^* + D^* G ) \left(4
   M_\nu^2+M_V^2\right) \left(M_V^2-4 M_\nu^2\right)^2
   \left(M_{Z^\prime}^2-4 M_\nu^2\right)^2}{M_{Z^\prime}^4} \nonumber\\
&&+\frac{| D |^2
   \left(M_V^2-4 M_\nu^2\right)^2 \left(M_{Z^\prime}^2-4
   M_\nu^2\right)^2}{M_{Z^\prime}^4} \nonumber\\
&&-\frac{(E F^* + E^* F ) \left(4
   M_\nu^2-M_V^2\right) \left(M_V^2-4 M_\nu^2\right)^2
   \left(M_{Z^\prime}^2-4 M_\nu^2\right)^2}{M_{Z^\prime}^4} \nonumber\\ 
&&-\frac{(E G^* + E^* G )
   \left(4 M_\nu^2+M_V^2\right) \left(M_V^2-4 M_\nu^2\right)^2
   \left(M_{Z^\prime}^2-4 M_\nu^2\right)^2}{M_{Z^\prime}^4} \nonumber\\
&& +\frac{| E |^2
   \left(M_V^2-4 M_\nu^2\right)^2 \left(M_{Z^\prime}^2-4
   M_\nu^2\right)^2}{M_{Z^\prime}^4} \nonumber\\
&& +\frac{(F G^* + F^* G ) \left(4
   M_\nu^2-M_V^2\right)^3 \left(4 M_\nu^2+M_V^2\right)
   \left(M_{Z^\prime}^2-4 M_\nu^2\right)^2}{2 M_{Z^\prime}^4} \nonumber\\
&&+\frac{| F |^2
   \left(M_V^2-4 M_\nu^2\right)^4 \left(M_{Z^\prime}^2-4 M_\nu^2\right)^2}{4
   M_{Z^\prime}^4} \nonumber\\
&& +\frac{| G |^2 \left(4 M_\nu^2+M_V^2\right)^2
   \left(M_V^2-4 M_\nu^2\right)^2 \left(M_{Z^\prime}^2-4 M_\nu^2\right)^2}{4
   M_{Z^\prime}^4}\biggr) \biggr]\,,
\nonumber\\
\nonumber\\
2 Re \left| M_\Phi \cdot M_{Z^\prime}^* \right| &=& Re \biggl[ \frac{a_{\nu\bar{\nu}}^{Z^\prime} y_{p,\nu\bar{\nu}}^\Phi M  \left(M_V^2-4 M^2\right)^2
   \left(4 M^2-M_{Z^\prime}^2\right)}{M_{Z^\prime}^2 \Sigma_\Phi \Sigma_{Z^\prime}^*} \times \nonumber\\
&& \,\,\,\,\,\,\,\,\,\,\,\,\,\,\,\, C \left(4 M^2
   (F^*+G^*)+M_V^2 (G^*-F^*)+2 D^*-2
   E^*\right) \biggr] \,.
\end{eqnarray}

%
%
\section{ One-loop annihilation into $V V$}
\label{app:V-V}

The annihilation channel $\nu \bar{\nu} \to V_1 V_2$ (where possible final states include $ZZ, Z^\prime Z^\prime, Z Z^\prime$ and $W^+ W^-$) gets contributions from both $s$-channel $\Phi$ and $Z^\prime$ exchange.  The amplitude from the $\Phi$-exchange diagram takes the form:
\begin{equation}
{\cal{V}}_{\Phi \to V_1 V_2}^{\mu \nu} = A g^{\mu \nu} + B p_2^\mu p_1^\nu + i C \epsilon^{p_1 p_2 \mu \nu} \, , 
\end{equation}
while the effective vertex for the $Z^\prime$-exchange diagram is given by:
\begin{equation}
{\cal{V}}_{Z^\prime \to V \gamma}^{\alpha \mu \nu} = D \epsilon^{p_1 \mu \nu \alpha} + E \epsilon^{p_2 \mu \nu \alpha} + F p_1^\alpha \epsilon^{p_1 p_2 \mu \nu} + G p_2^\alpha \epsilon^{p_1 p_2 \mu \nu} + H p_1^\nu \epsilon^{p_1 p_2 \alpha \mu} + I p_2^\mu \epsilon^{p_1 p_2 \alpha \nu} \,.
\end{equation}

The loop coefficients for the $\Phi$-exchange diagram are (for simplicity, we assume $V_1 = V_2$):
\begin{eqnarray}
A &=& -\frac{i y_{s,f\bar{f}}^{Z^\prime} m_f}{8 \pi ^2
   \left(M_\nu^2-M_V^2\right)} \biggl(\left( a_{f\bar{f}}^V \right)^2 \biggl(M_V^2
   \biggl(3 B_0(12)+2 B_0(13)+3 B_0(23)-2 C_0
   M_V^2 \nonumber \\
&&+8 C_0 m_f^2-4\biggr)-4 M_\nu^2
   \left(B_0(12)+B_0(23)+C_0 M_V^2+2 C_0
   m_f^2-1\right)+8 C_0 M_\nu^4\biggr) \nonumber\\
&&+\left( v_{f\bar{f}}^V \right)^2
   \biggl(-M_V^2 \left(B_0(12)-2 B_0(13)+B_0(23)+2
   C_0 M_V^2+8 C_0 m_f^2+4\right) \nonumber\\
&&-8 C_0
   M_\nu^4+4 M_\nu^2 \left(3 C_0 M_V^2+2 C_0
   m_f^2+1\right)\biggr)\biggr)\,, \\
\nonumber\\
B &=& \frac{i y_{s,f\bar{f}}^{Z^\prime} m_f}{16 \pi ^2 M^2
   \left(M^2-M_V^2\right)^2} \biggl(\left( a_{f\bar{f}}^{V} \right)^2 \left(2
   M^2-M_V^2\right) \biggl(M^2 \biggl(4 B_0(13)-4 B_0(23)-2
   C_0 M_V^2 \nonumber\\
&&+4 C_0 m_f^2+2\biggr)+M_V^2
   \left(-B_0(13)+B_0(23)+C_0 M_V^2-4 C_0
   m_f^2-2\right)+4 C_0 M^4\biggr) \nonumber\\
&&+\left( v_{f\bar{f}}^{V} \right)^2 \biggl(-2 M^2
   M_V^2 \left(-B_0(13)+B_0(23)+2 C_0 M_V^2+6
   C_0 m_f^2+3\right) \nonumber\\
&&+M_V^4
   \left(B_0(13)-B_0(23)-C_0 M_V^2+4 C_0
   m_f^2+2\right)-8 C_0 M^6 \nonumber\\
&&+4 M^4 \left(2 C_0 \left(2
   M_V^2+m_f^2\right)+1\right)\biggr)\biggr) \,, \\
\nonumber\\
C &=& -\frac{y_{p,f\bar{f}}^{Z^\prime} m_f}{2
   \pi ^2 (M-M_V) (M+M_V)} \biggl(B_0(13) \left( a_{f\bar{f}}^{V} \right)^2-B_0(23)
   \left( a_{f\bar{f}}^{V} \right)^2 \nonumber\\
&&+C_0 M^2 (a_{f\bar{f}}^{V}-v_{f\bar{f}}^{V})
   (a_{f\bar{f}}^{V}+v_{f\bar{f}}^{V})+C_0 \left( v_{f\bar{f}}^{V} \right)^2 M_V^2\biggr) \,,
\end{eqnarray}
while the loop coefficients for the $Z^\prime$-exchange diagram are:
\begin{eqnarray}
D &=& \frac{1}{4\pi^2} \left(a_{f\bar{f}}^{Z^\prime}
   \left(\left( a_{f\bar{f}}^{V} \right)^2+\left( v_{f\bar{f}}^{V} \right)^2\right)+2 a_{f\bar{f}}^{V} v_{f\bar{f}}^{V}
   v_{f\bar{f}}^{Z^\prime}\right)-2 a_{f\bar{f}}^{Z^\prime} m_f^2 \biggl(B_0(13)
   \left( a_{f\bar{f}}^{V} \right)^2 \nonumber\\
&&-B_0(23) \left( a_{f\bar{f}}^{V} \right)^2+C_0 M_\nu^2
   (a_{f\bar{f}}^{V}-v_{f\bar{f}}^{V}) (a_{f\bar{f}}^{V}+v_{f\bar{f}}^{V})+C_0
   \left( v_{f\bar{f}}^{V} \right)^2 M_V^2\biggr)\,, \\
\nonumber\\
E &=& \frac{2 a_{f\bar{f}}^{Z^\prime} m_f^2}{4 \pi ^2 (M_\nu^2-M_V^2)} \biggl(B_0(13) \left( a_{f\bar{f}}^V \right)^2-B_0(23)
   \left( a_{f\bar{f}}^V \right)^2+C_0 M_\nu^2 (a_{f\bar{f}}^V-v_{f\bar{f}}^V)
   (a_{f\bar{f}}^V+v_{f\bar{f}}^V) \nonumber\\
&&+C_0 \left( v_{f\bar{f}}^V \right)^2
   M_V^2\biggr)-(M_\nu-M_V) (M_\nu+M_V) \left(a_{f\bar{f}}^{Z^\prime}
   \left(\left( a_{f\bar{f}}^V \right)^2+\left( v_{f\bar{f}}^V \right)^2\right)+2 a_{f\bar{f}}^V v_{f\bar{f}}^V
   v_{f\bar{f}}^{Z^\prime}\right) \,, \\
\nonumber\\
F &=& 0 \,, \\
\nonumber\\
G &=& 0 \,, \\
\nonumber\\
H &=& -\frac{\left(a_{f\bar{f}}^{Z^\prime} \left(\left( a_{f\bar{f}}^{V} \right)^2+\left( v_{f\bar{f}}^{V} \right)^2\right)+2
   a_{f\bar{f}}^{V} v_{f\bar{f}}^{V} v_{f\bar{f}}^{Z^\prime}\right)}{32 \pi^2 M_\nu^2
   \left(M_\nu^2-M_V^2\right)^2} \biggl(2 M_\nu^2 M_V^2
   \biggl(B_0(13)-B_0(23)+2 C_0 M_V^2 \nonumber\\
&& -6 C_0
   m_f^2-3\biggr)+M_V^4
   \left(B_0(13)-B_0(23)-C_0 M_V^2+4 C_0
   m_f^2+2\right) \nonumber\\
&& +M_\nu^4 \left(8 C_0
   m_f^2+4\right)\biggr) \,, \\
\nonumber\\
I &=& \frac{\left(a_{f\bar{f}}^{Z^\prime} \left(\left( a_{f\bar{f}}^{V} \right)^2+\left( v_{f\bar{f}}^{V} \right)^2\right)+2
   a_{f\bar{f}}^{V} v_{f\bar{f}}^{V} v_{f\bar{f}}^{Z^\prime}\right)}{32 \pi^2 M_\nu^2
   \left(M_\nu^2-M_V^2\right)^2} \biggl(2 M_\nu^2 M_V^2
   \biggl(B_0(13)-B_0(23)+2 C_0 M_V^2 \nonumber\\
&& -6 C_0
   m_f^2-3\biggr)+M_V^4
   \left(B_0(13)-B_0(23)-C_0 M_V^2+4 C_0
   m_f^2+2\right) \nonumber\\
&& +M_\nu^4 \left(8 C_0
   m_f^2+4\right)\biggr) \,.
\end{eqnarray}

The contributions to the total amplitude-squared are:
\begin{eqnarray}
\left| {\cal M}_\Phi \right|^2 &=& \frac{8 \left( y_{p, \nu\bar{\nu}}^\Phi \right)^2 M^2}{M_V^4 | \Sigma_\Phi |^2} \biggl(8 M^2 (M-M_V) (M+M_V)
   \biggl(-M_V^2 \left((A B^* + A^* B)+2 |B|^2 M^2\right) \nonumber\\
&&+2
   (A B^* + A^* B) M^2+2 |B|^2 M^4+|C|^2
   M_V^4\biggr) \nonumber\\
&& +|A|^2 \left(4 M^4-4 M^2 M_V^2+3
   M_V^4\right)\biggr) \,,
\end{eqnarray}
\begin{eqnarray}
\left| {\cal M}_{Z^\prime}\right|^2 &=& \frac{16}{M_V^2 M_{Z^\prime}^4 |\Gamma_{Z^\prime}|^2} \biggl(M_\nu^2 M_{Z^\prime}^4 \biggl(M_V^4
   \biggl(\left(\left( a_{\nu\bar{\nu}}^{Z^\prime} \right)^2-2 \left( v_{\nu\bar{\nu}}^{Z^\prime} \right)^2\right) \biggl(2
   (D E^* + D^* E) \nonumber\\
&&-|D|^2-|E|^2 \biggr)+8 \left( v_{\nu\bar{\nu}}^{Z^\prime} \right)^2 M_\nu^2
   ((D H^* + D^* H)-(D I^* + D^* I) \nonumber\\
&&-(E H^* + E^* H)+(E I^* + E^* I))+16 \left( v_{\nu\bar{\nu}}^{Z^\prime} \right)^2 M_\nu^4
   (|H|^2+|I|^2)\biggr) \nonumber\\
&&+M_\nu^2 M_V^2 \biggl(-2 (D E^* + D^* E)
   \left(\left( a_{\nu\bar{\nu}}^{Z^\prime} \right)^2-5 \left( v_{\nu\bar{\nu}}^{Z^\prime} \right)^2\right) \nonumber\\
&& +8 \left( v_{\nu\bar{\nu}}^{Z^\prime} \right)^2 M_\nu^2 (-3
   (D H^* + D^* H)+(D I^* + D^* I)+(E H^* + E^* H) \nonumber\\
&& -3
   (E I^* + E^* I))+(|D|^2+|E|^2) \left(\left( a_{\nu\bar{\nu}}^{Z^\prime} \right)^2-3
   \left( v_{\nu\bar{\nu}}^{Z^\prime} \right)^2\right) \nonumber\\
&&-32 \left( v_{\nu\bar{\nu}}^{Z^\prime} \right)^2 M_\nu^4
   (|H|^2+|I|^2)\biggr)+4 \left( v_{\nu\bar{\nu}}^{Z^\prime} \right)^2 M_\nu^4 \biggl(4 M_\nu^2
   \biggl((D H^* + D^* H) \nonumber\\
&& +(E I^* + E^* I)+M_\nu^2
   (|H|^2+|I|^2)\biggr)+|D|^2+|E|^2\biggr)\biggr)
   \nonumber\\
&&+16 \left( a_{\nu\bar{\nu}}^{Z^\prime} \right)^2 M_\nu^6 M_V^2 (M_\nu -M_V) (M_\nu +M_V) (-2
   (D E^* + D^* E)+|D|^2+|E|^2) \nonumber\\
&&-8 \left( a_{\nu\bar{\nu}}^{Z^\prime} \right)^2 M_\nu^4 M_V^2
   M_{Z^\prime}^2 (M_\nu -M_V) (M_\nu +M_V) (-2
   (D E^* + D^* E) \nonumber\\
&& +|D|^2+|E|^2)\biggr)
\end{eqnarray} 
\begin{eqnarray}
2 Re \left| M_\Phi \cdot M_{Z^\prime}^* \right| &=& Re \biggl[ \frac{32 a_{\nu\bar{\nu}}^{Z^\prime} y_{p,\nu\bar{\nu}}^\Phi M_\nu^3 C (M_\nu^2-M_V^2)
   \left(4 M_\nu^2-M_{Z^\prime}^2\right) (D^*-E^*)}{M_{Z^\prime}^2 \Gamma_\Phi \left( \Gamma_{Z^\prime} \right)^*} \biggr]
\end{eqnarray}																																							
%
%
\section{ One-loop annihilation into $\gamma \Phi$}
\label{app:gamma-Phi}

The annihilation mode $\nu \bar{\nu} \to \gamma \Phi$ can only proceed via $s$-channel $Z^\prime$ exchange.  The effective vertex for the $Z^\prime \gamma \Phi$ interaction takes the form:
\begin{equation}
{\cal{V}}_{Z^\prime \to \gamma \Phi}^{\alpha \mu} = A g^{\alpha\mu} + i B \epsilon^{p_1 p_2 \alpha \mu} \,,
\end{equation}
where the loop coefficients are:
\begin{eqnarray}
A &=& -\frac{v_{f\bar{f}}^\gamma y_{s, f\bar{f}}^\Phi v_{f\bar{f}}^{Z^\prime} m_f}{4 \pi ^2
   \biggl(4 M_\nu^2-M_\Phi^2\biggr)} \biggl(8 M_\nu^2
   \biggl(B_0(13)-B_0(23)-C_0 M_\Phi^2+2 C_0
   m_f^2+1\biggr) \nonumber\\
&&\,\,\,\,\,\,\,\,\,\,\,\,\,\,\,\,\,\, +16 C_0 M_\nu^4 +M_\Phi^2 \biggl(C_0
   \biggl(M_\Phi^2-4 m_f^2\biggr)-2\biggr)\biggr) \,, \\
\nonumber\\
B &=& -\frac{v_{f\bar{f}}^\gamma y_{p, f\bar{f}}^\Phi v_{f\bar{f}}^{Z^\prime} m_f C_0}{2 \pi ^2} \,.
\end{eqnarray}

The amplitude-squared is:
\begin{equation}
\left| {\cal M} \right|^2 = \frac{v_{\nu\bar{\nu}}^2}{4 \left| \Sigma_{Z^\prime} \right|^2} \left[ 4 \left| A \right|^2 + \left(4 M_\nu^2 - M_\Phi^2\right)^2 \left| B \right|^2 \right]
\end{equation}

%
%
\section{ One-loop annihilation into $V \Phi$}
\label{app:V-Phi}

As for the $\gamma \Phi$ channel, the annihilation into $V\Phi$ (where $V = Z, Z^\prime$) can only proceed via an $s$-channel $Z^\prime$.  The effective vertex is given by:
\begin{equation}
{\cal{V}}_{Z^\prime \to V \Phi}^{\alpha \mu} = A g^{\alpha \mu} + B p_1^\alpha p_2^\mu + C p_1^\alpha p_2^\mu + i D \epsilon^{p_1 p_2 \mu \alpha} \,.
\end{equation}
The loop coefficients for this process are:
\begin{eqnarray}
A &=& -\frac{v_{f\bar{f}}^{V} m_f}{4 \pi ^2
   \biggl(-2 M_\Phi^2 \biggl(4
   M_\nu^2+M_V^2\biggr)+\biggl(M_V^2-4
   M_\nu^2\biggr)^2+M_\Phi^4\biggr)} \times \nonumber\\
&& \biggl(y_{p, f\bar{f}}^\Phi a_{f\bar{f}}^{Z^\prime} \biggl(-4 M_\nu^2
   \biggl(-2 M_V^2 \biggl(B_0(12)+B_0(13)+2
   B_0(23)-C_0 M_\Phi^2+4 C_0
   m_f^2-2\biggr) \nonumber\\
&&+M_\Phi^2 \biggl(-4 B_0(12)-2 B_0(13)-2
   B_0(23)+C_0 M_\Phi^2-8 C_0
   m_f^2+4\biggr)+C_0
   M_V^4\biggr) \nonumber\\
&&+(M_V-M_\Phi) (M_V+M_\Phi)
   \biggl(M_\Phi^2 \biggl(2 B_0(12)+2 B_0(13)-C_0
   M_\Phi^2+4 C_0 m_f^2-2\biggr) \nonumber\\
&& -2 M_V^2
   \biggl(B_0(13)+B_0(23)+2 C_0
   m_f^2-1\biggr)+C_0 M_V^4\biggr)-16 M_\nu^4 \biggl(2
   B_0(12)+2 B_0(23) \nonumber\\
&&+C_0 \biggl(M_V^2+M_\Phi^2+4
   m_f^2\biggr)-2\biggr)+64 C_0 M_\nu^6\biggr)+y_{s, f\bar{f}}^\Phi
   v_{f\bar{f}}^{Z^\prime} \biggl(-4 M_\nu^2 \biggl(2 M_V^2
   \biggl(B_0(12) \nonumber\\
&&+B_0(13)-2 B_0(23)-C_0 M_\Phi^2+4
   C_0 m_f^2+2\biggr)+M_\Phi^2 \biggl(2 B_0(13)-2
   B_0(23)-3 C_0 M_\Phi^2 \nonumber\\
&&+8 C_0
   m_f^2+4\biggr)+C_0
   M_V^4\biggr)+(M_V-M_\Phi) (M_V+M_\Phi) \biggl(2
   M_V^2 \biggl(B_0(12)-B_0(23) \nonumber\\
&&-C_0 M_\Phi^2+2
   C_0 m_f^2+1\biggr)+C_0 M_V^4+M_\Phi^2
   \biggl(C_0 M_\Phi^2-4 C_0
   m_f^2-2\biggr)\biggr) \nonumber\\
&&+16 M_\nu^4 \biggl(2 B_0(13)-2
   B_0(23)-C_0 \biggl(M_V^2+3 M_\Phi^2-4
   m_f^2\biggr)+2\biggr)+64 C_0 M_\nu^6\biggr)\biggr) \, , \\
  \nonumber
\end{eqnarray}
\begin{eqnarray}
B &=& -\frac{a_{f\bar{f}}^V M_V^2 m_f (y_{p,f\bar{f}}^\Phi a_{f\bar{f}}^{Z^\prime}+y_{s,f\bar{f}}^\Phi
   v_{f\bar{f}}^{Z^\prime})}{\pi ^2 \biggl(-2 M_\Phi^2 \biggl(4
   M_\nu^2+M_V^2\biggr)+\biggl(M_V^2-4
   M_\nu^2\biggr)^2+M_\Phi^4\biggr)^2}  \times \nonumber\\
&&\biggl(-16 M_\nu^4 \biggl(2 B_0(12)-4 B_0(13)+2
   B_0(23)+C_0 \biggl(M_V^2+M_\Phi^2-4
   m_f^2\biggr)-2\biggr) \nonumber\\
&&-4 M_\nu^2 \biggl(2 M_V^2
   \biggl(B_0(12)+B_0(13)-2 B_0(23)-3 C_0 M_\Phi^2+4
   C_0 m_f^2+2\biggr) \nonumber\\
&&+M_\Phi^2 \biggl(-4 B_0(12)+2
   B_0(13)+2 B_0(23)+C_0 M_\Phi^2+8 C_0
   m_f^2+4\biggr)+C_0
   M_V^4\biggr) \nonumber\\
&&+(M_V-M_\Phi) (M_V+M_\Phi)
   \biggl(M_V^2 \biggl(4 B_0(12)-2 B_0(13)-2 B_0(23)+4
   C_0 m_f^2+2\biggr) \nonumber\\
&&-M_\Phi^2 \biggl(-2 B_0(12)-2
   B_0(13)+4 B_0(23)+C_0 M_\Phi^2+4 C_0
   m_f^2+2\biggr)+C_0 M_V^4\biggr) \nonumber\\ 
&&+64 C_0
   M_\nu^6\biggr) \, , \\
  \nonumber
\end{eqnarray}
\begin{eqnarray}
C &=& \frac{v_{f\bar{f}}^V m_f}{2 \pi ^2 (2 M_\nu -M_V+M_\Phi)^2 (2
   M_\nu +M_V+M_\Phi)^2 \biggl(M_V^2-(M_\Phi-2 M)^2\biggr)^2} \times \nonumber\\
&& \biggl(y_{p, f\bar{f}}^\Phi a_{f\bar{f}}^{Z^\prime} \biggl(4
   M_\nu^2-M_V^2-M_\Phi^2\biggr) \biggl(-16 M_\nu^4 \biggl(2 B_0(12)-4
   B_0(13)+2 B_0(23) \nonumber\\
&&+C_0 \biggl(M_V^2+M_\Phi^2-4
   m_f^2\biggr)-2\biggr)-4 M_\nu^2 \biggl(2 M_V^2
   \biggl(B_0(12)+B_0(13)-2 B_0(23) \nonumber\\
&&-3 C_0 M_\Phi^2+4
   C_0 m_f^2+2\biggr)+M_\Phi^2 \biggl(-4 B_0(12)+2
   B_0(13)+2 B_0(23)+C_0 M_\Phi^2+8 C_0
   m_f^2+4\biggr) \nonumber\\ 
&&+C_0
   M_V^4\biggr)+(M_V-M_\Phi) (M_V+M_\Phi)
   \biggl(M_V^2 \biggl(4 B_0(12)-2 B_0(13)-2 B_0(23)+4
   C_0 m_f^2+2\biggr) \nonumber\\
&&-M_\Phi^2 \biggl(-2 B_0(12)-2
   B_0(13)+4 B_0(23)+C_0 M_\Phi^2+4 C_0
   m_f^2+2\biggr)+C_0 M_V^4\biggr)+64 C_0
   M_\nu^6\biggr) \nonumber\\
&+& y_{s, f\bar{f}}^\Phi v_{f\bar{f}}^{Z^\prime} \biggl(-32 M_\nu^4 \biggl(M_V^2
   \biggl(3 B_0(12)-3 B_0(23)-5 C_0 M_\Phi^2+6
   C_0 m_f^2+3\biggr)+M_\Phi^2 \biggl(2 B_0(13) \nonumber\\
&&-2
   B_0(23)-3 C_0 M_\Phi^2+6 C_0
   m_f^2+3\biggr)\biggr)+8 M_\nu^2 \biggl(M_V^4 \biggl(6
   B_0(12)-3 B_0(13)-3 B_0(23) \nonumber\\
&&-2 C_0 M_\Phi^2+6
   C_0 m_f^2+3\biggr)+M_V^2 M_\Phi^2 \biggl(6
   B_0(13)-6 B_0(23)-C_0 M_\Phi^2+4 C_0
   m_f^2+2\biggr) \nonumber\\
&&+M_\Phi^4 \biggl(B_0(13)-B_0(23)-2
   C_0 M_\Phi^2+6 C_0 m_f^2+3\biggr)+C_0
   M_V^6\biggr) \nonumber\\
&&-(M_V-M_\Phi) (M_V+M_\Phi)
   \biggl(M_V^4 \biggl(6 B_0(12)-4 B_0(13)-2 B_0(23)+3
   C_0 M_\Phi^2+4 C_0 m_f^2+2\biggr) \nonumber\\
&&+M_V^2
   M_\Phi^2 \biggl(6 B_0(12)+4 B_0(13)-10 B_0(23)-5
   C_0 M_\Phi^2\biggr)+C_0 M_V^6+M_\Phi^4
   \biggl(C_0 M_\Phi^2-4 C_0
   m_f^2 \nonumber\\
&&-2\biggr)\biggr)+128 M_\nu^6
   \biggl(B_0(13)-B_0(23)-C_0 \biggl(M_V^2+2
   M_\Phi^2-2 m_f^2\biggr)+1\biggr)+256 C_0
   M_\nu^8\biggr)\biggr) \, , \\
  \nonumber
\end{eqnarray}
\begin{eqnarray}
D &=& -\frac{a_{f\bar{f}}^V m_f}{2 \pi ^2 \biggl(-2
   M_\Phi^2 \biggl(4 M_\nu^2+M_V^2\biggr)+\biggl(M_V^2-4
   M_\nu^2\biggr)^2+M_\Phi^4\biggr)} \times   \nonumber \\
&&\biggl(2 M_\Phi^2 \biggl(y_{p, f\bar{f}}^\Phi
   a_{f\bar{f}}^{Z^\prime} (B_0(12)+B_0(13)-2 B_0(23))-C_0
   y_{s, f\bar{f}}^\Phi v_{f\bar{f}}^{Z^\prime} \biggl(4 M_\nu^2+M_V^2\biggr)\biggr) \nonumber\\
&&+\biggl(4
   M_\nu^2-M_V^2\biggr) \biggl(-2 B_0(12) y_{p, f\bar{f}}^\Phi a_{f\bar{f}}^{Z^\prime}+2
   B_0(13) y_{p, f\bar{f}}^\Phi a_{f\bar{f}}^{Z^\prime} \nonumber\\
&&+C_0 \biggl(4
   M_\nu^2-M_V^2\biggr) (y_{p, f\bar{f}}^\Phi a_{f\bar{f}}^{Z^\prime}+y_{s, f\bar{f}}^\Phi
   v_{f\bar{f}}^{Z^\prime})\biggr) +C_0 M_\Phi^4 (y_{s, f\bar{f}}^\Phi
   v_{f\bar{f}}^{Z^\prime}-y_{p, f\bar{f}}^\Phi a_{f\bar{f}}^{Z^\prime})\biggr)
\end{eqnarray}

Finally, the matrix-element-squared is given by:
\begin{eqnarray}
\left| {\cal M} \right|^2 &=& \frac{1}{\left| \Sigma_{Z^\prime} \right|^2} \biggl[
\frac{( A B^* + A^* B)}{16
   M_V^2 M_{Z^\prime}^4} \biggl(-2 M_\Phi^2 \biggl(4
   M^2+M_V^2\biggr)+\biggl(M_V^2-4
   M^2\biggr)^2+M_\Phi^4\biggr) \times \nonumber\\
&& \biggl(\left( a_{\nu\bar{\nu}}^{Z^\prime} \right)^2 \biggl(M_{Z^\prime}^2-4
   M^2\biggr)^2 \biggl(4 M^2-M_V^2+M_\Phi^2\biggr)+\left( v_{\nu\bar{\nu}}^{Z^\prime} \right)^2
   M_{Z^\prime}^4 \biggl(-4 M^2-M_V^2+M_\Phi^2\biggr)\biggr) \nonumber\\
&+&\frac{( A C^* + A^* C)}{16 M_V^2 M_{Z^\prime}^4} (-2 M_\nu +M_V-M_\Phi)
   (2 M_\nu +M_V-M_\Phi) (-2 M_\nu +M_V+M_\Phi)  \times \nonumber\\
&&(2
   M_\nu +M_V+M_\Phi) \biggl(4 M^2+M_V^2-M_\Phi^2\biggr)
   \biggl(\left( a_{\nu\bar{\nu}}^{Z^\prime} \right)^2 \biggl(M_{Z^\prime}^2-4 M^2\biggr)^2 \nonumber\\
&&+\left( v_{\nu\bar{\nu}}^{Z^\prime} \right)^2
   M_{Z^\prime}^4\biggr){16 M_V^2 M_{Z^\prime}^4} \nonumber\\
&+&\frac{| A |^2}{8 M_V^2 M_{Z^\prime}^4}
   \biggl(\biggr( a_{\nu\bar{\nu}}^{Z^\prime} \biggr)^2 \biggl(M_{Z^\prime}^2-4 M^2\biggr)^2 \biggl(-2
   M_\Phi^2 \biggl(4 M^2+M_V^2\biggr)+\biggl(M_V^2-4
   M^2\biggr)^2+M_\Phi^4\biggr) \nonumber\\
&&+\left( v_{\nu\bar{\nu}}^{Z^\prime} \right)^2 M_{Z^\prime}^4 \biggl(16
   M^4+8 M^2 \biggl(5
   M_V^2-M_\Phi^2\biggr)+\biggl(M_V^2-M_\Phi^2\biggr)^2\biggr)\biggr) \nonumber\\
&+& \frac{( B C^* + B^* C)}{32 M_V^2
   M_{Z^\prime}^4} \biggl(-2
   M_\Phi^2 \biggl(4 M^2+M_V^2\biggr)+\biggl(M_V^2-4
   M^2\biggr)^2+M_\Phi^4\biggr) \times \nonumber\\
&& \biggl(\left( a_{\nu\bar{\nu}}^{Z^\prime} \right)^2 \biggl(M_{Z^\prime}^2-4
   M^2\biggr)^2 \biggl(16
   M^4-\biggl(M_V^2-M_\Phi^2\biggr)^2\biggr) \nonumber\\
&&-\left( v_{\nu\bar{\nu}}^{Z^\prime} \right)^2
   M_{Z^\prime}^4 \biggl(-2 M_\Phi^2 \biggl(4
   M^2+M_V^2\biggr)+\biggl(M_V^2-4
   M^2\biggr)^2+M_\Phi^4\biggr)\biggr) \nonumber\\
&+& \frac{| B |^2}{32 M_V^2
   M_{Z^\prime}^4} \biggl(-2 M_\Phi^2 \biggl(4
   M^2+M_V^2\biggr)+\biggl(M_V^2-4
   M^2\biggr)^2+M_\Phi^4\biggr) \times \nonumber \\
&& \biggl(\left( a_{\nu\bar{\nu}}^{Z^\prime} \right)^2 \biggl(M_{Z^\prime}^2-4
   M^2\biggr)^2 \biggl(4 M^2-M_V^2+M_\Phi^2\biggr)^2 \nonumber\\
&&+\left( v_{\nu\bar{\nu}}^{Z^\prime} \right)^2
   M_{Z^\prime}^4 \biggl(-2 M_\Phi^2 \biggl(4
   M^2+M_V^2\biggr)+\biggl(M_V^2-4
   M^2\biggr)^2+M_\Phi^4\biggr)\biggr) \nonumber\\
&+& \frac{| C |^2}{32 M_V^2
   M_{Z^\prime}^4} \biggl(-2 M_\Phi^2 \biggl(4
   M^2+M_V^2\biggr)+\biggl(M_V^2-4
   M^2\biggr)^2+M_\Phi^4\biggr) \times \nonumber\\
&& \biggl(\left( a_{\nu\bar{\nu}}^{Z^\prime} \right)^2 \biggl(M_{Z^\prime}^2-4
   M^2\biggr)^2 \biggl(4 M^2+M_V^2-M_\Phi^2\biggr)^2 \nonumber\\
&&+\left( v_{\nu\bar{\nu}}^{Z^\prime} \right)^2
   M_{Z^\prime}^4 \biggl(-2 M_\Phi^2 \biggl(4
   M^2+M_V^2\biggr)+\biggl(M_V^2-4
   M^2\biggr)^2+M_\Phi^4\biggr)\biggr) \nonumber\\
&+&| D |^2 \left( v_{\nu\bar{\nu}}^{Z^\prime} \right)^2 M^2 \biggl(-2 M_\Phi^2 \biggl(4
   M^2+M_V^2\biggr)+\biggl(M_V^2-4
   M^2\biggr)^2+M_\Phi^4\biggr) \, .
\biggr]
\end{eqnarray}


\end{document}